\documentclass[11pt,final,3p,times]{elsarticle}

\usepackage{amssymb}
\usepackage{amsmath}
\usepackage{lineno}
\usepackage{wasysym}
\usepackage{xcolor}
\usepackage{txfonts}
\usepackage{url}

\newcommand{\near}{\text{near}}
\newcommand{\far}{\text{far}}

\newcommand{\twofigures}[2]{
  \begin{figure}[htbp]
    \begin{minipage}[b]{0.49\linewidth}
      #1
    \end{minipage}
    \hfill
    \begin{minipage}[b]{0.49\linewidth}
      #2
    \end{minipage}
  \end{figure}
}

\journal{Nuclear Instruments and Methods A}

\begin{document}

\makeatletter
\def\ps@pprintTitle{%
 \let\@oddhead\@empty
 \let\@evenhead\@empty
 \def\@oddfoot{\centerline{\thepage}}%
 \let\@evenfoot\@oddfoot}
\makeatother

\begin{frontmatter}

\title{Alignment-related Effects in Forward Proton Experiments at the LHC
}

\author[]{R. Staszewski\corref{cor1}}

\cortext[cor1]{Corresponding author}

\ead{Rafal.Staszewski@ifj.edu.pl}

\author[]{J.~Chwastowski}
\ead{Janusz.Chwastowski@ifj.edu.pl}

\author[]{K. Korcyl}
\ead{Krzysztof.Korcyl@ifj.edu.pl}

\author[]{M. Trzebi\'nski}
\ead{Maciej.Trzebinski@ifj.edu.pl}
\address{The Henryk Niewodnicza\'nski Institute of Nuclear Physics\\ Polish
  Academy of Sciences,\\ ul. Radzikowskiego 152, 31-342 Krak\'ow, Poland
}

\begin{abstract}
  
  The activity in the field of diffractive physics at the Large Hadron Collider
  has been constantly increasing. This includes the planning for additional
  dedicated apparatus -- horizontal forward proton detectors. This work
  focuses on the problems related to the alignment of such devices. The
  effects of the misalignment of the detectors on their geometric acceptance
  and on the reconstruction of the proton kinematics are studied. The requirements for the
  alignment precision are inferred for different types of possible measurements.

\end{abstract}


\end{frontmatter}



\section{Introduction}

Studies of Standard Model physics are an important part of the Large Hadron Collider experimental programme.
One of the least understood branches of the Standard Model is diffraction and, in particular, hard diffractive processes.

Diffraction in soft hadron collisions is a well-established phenomenon.
The existence of hard diffractive interactions were confirmed for the first time at the CERN SPS collider in the UA8 experiment.
Later, the HERA experiments, H1 and ZEUS, measured a significant contribution of diffraction to many processes, including DIS.
This was followed by another interesting observation at the Tevatron,
where the extrapolations based on HERA measurements led to the overestimation of the hard diffractive cross section by a factor of about 10.
Although today these issues are considered to be reasonably well understood,
the experience gained shows that 
agreement of LHC results with expectations should not be taken for granted.

Two properties are typically used to discriminate between diffractive and non-diffractive interactions.
In non-diffractive events colour charge is exchanged between the interacting protons%
\footnote{Although diffractive interactions are possible in any hadron-hadron collision, in this work we focus on proton-proton interactions at the LHC.},
leading to the break-up of the protons and to enhanced production of hadrons in the forward region.
By contrast, in the case of diffractive interactions a colour singlet is exchanged between the colliding protons, and forward production of particles is suppressed.
As a result, the rapidity distribution of final state particles in diffractive events contains regions completely devoid of particles -- the so-called large rapidity gaps.

The size of the rapidity gap is anti-correlated to the energy transferred from the diffractively scattered proton and, in consequence, to the diffractive mass.
Moreover, rapidity gaps can be present also in non-diffractive events, where they can emerge due to fluctuations of the distance between neighbouring particles.
The size of such a gap has a steeply falling exponential distribution.
The size of the gap in diffractive events drops more slowly, which makes the rapidity gap selection method appropriate for events with large gaps, \textit{i.e.} small diffractive masses.
For diffractive processes with large masses in the final state the use of this method is problematic.
In addition, the high pile-up environment of the LHC makes it impossible for calorimeters to be used in the rapidity gap reconstruction.

The second method used to measure diffraction is based on the fact that the exchange of a colour singlet may leave the proton intact.
Such a proton is characterised by a very steep distribution of the scattering angle.
In fact, at the LHC, the scattered diffractive protons remain inside the beam pipe and traverse the magnetic structures of the accelerator accompanying the proton bunch.
However, the kinematics of the diffractive protons is slightly different from that of the protons of the beam (greater transverse momentum or lower longitudinal momentum).
Because of this the trajectories of such protons recede from the beam orbit.
At a far enough distance from the interaction point (more than hundred meter), the diffractive protons may depart far enough from the beam core to be detected.

Measurements of diffractively scattered protons in the vicinity of the beam require dedicated detectors.
To access small scattering angles, the detectors have to be placed inside the accelerator beam pipe.
Moreover, it must be possible to adjust the distance between the detector and the beam during the operation.
This is because the minimal distance depends on the actual condition of the beam.
Typically, during beam injection and acceleration, when the beams are unstable, the detectors must be retracted into the parking positions.
This is possible by using a dedicated system -- the Roman pot \cite{Amaldi:1972uw} or the Hamburg movable beam pipe \cite{Schneekloth:1994,Albrow:2008pn}.
Depending on the nature of the studied process, the detectors can be designed to move in the machine plane (horizontal detectors) or along the normal to this plane (vertical detectors).

The procedures of inserting and retracting the detectors imply that the positions of the detectors will vary from one data taking period to another.
This makes the apparatus alignment more difficult than for other detectors, since it needs to be performed more frequently -- typically for each run.
In order to develop the alignment methods, one needs to know the precision needed to perform the physics measurement.
The answer to this problem is the subject of this paper.

\section{The AFP Detectors}

Presently at the LHC, the proton tagging detectors are installed at Point 1 (ALFA) \cite{alfa} and Point 5 (TOTEM) \cite{totem}.
The CMS and TOTEM collaborations plan to install additional horizontal detectors: 
recently, the CT-PPS (CMS-TOTEM Precision Proton Spectrometer) project \cite{Albrow:1753795} has been approved.
In addition to the existing vertical detectors around LHC Point 1, the installation of horizontal stations -- the AFP (ATLAS Forward Proton) detectors \cite{afp} -- is considered.
In this paper the alignment for the AFP detectors will be studied; however, the results for the TOTEM horizontal pots and CT-PPS should be similar.

The ultimate aim of the AFP detectors is to measure diffractive and two-photon processes during the periods when the LHC will work with standard settings, \textit{i.e.} low $\beta^\ast$, high luminosity.
It is planned to use horizontally inserted detectors placed symmetrically w.r.t.
the interaction point at 204 m (near station) and 212 m (far station).
The AFP stations will contain silicon pixel detectors \cite{Grinstein:2013nua} with the foreseen spatial resolution of 10 $\muup$m in the horizontal direction and 30 $\muup$m in the vertical direction, respectively.
The far stations will be also equipped with precise timing detectors with the resolution of 30 ps.

The protons scattered diffractively at the interaction point traverse  the fields of the LHC magnets together with the beam.
For the AFP detectors the magnets are: three quadrupole magnets (Q1, Q2, Q3 -- the inner triplet), two dipole magnets (D1 and D2 -- separating the incoming and outgoing beams) and two additional quadrupole magnets (Q4 and Q5).
The trajectory of a scattered proton depends on its momentum and the position of the interaction vertex.
Tracking of the proton through the magnetic lattice can be simulated with dedicated tools, such as the Mad-X \cite{mad} or the FPTrack \cite{fptracker} programs.

For the nominal LHC optics ($\beta^\ast = 0.55\,\text{m}$) with beam energy $p_0 = 7\,\text{TeV}$ and beam emittance of 3.5~$\muup$m$\cdot$rad, one can calculate the transverse size and the angular spread of the beam.
For example, at 212 metres from the interaction point the beam width equals 140 $\muup$m and 430 $\muup$m in the horizontal and vertical direction, respectively.
The information about the beam transverse size at the detector location is of particular interest.
This is because the detectors cannot approach the beam too closely, due to the radiation hazard for both the detectors themselves and the magnets located behind them, which is due to the hadronic showers generated in the Roman pot material.
The actual distance between the detector edge and the beam position depends on the beam intensity and the amount of the halo background.
A realistic distance value is between 10 and 20 times the width of the beam at the appropriate location ($\sigma$), depending on the intensity and condition of the beam.

The distance between the detector and the beam is of principal importance for the calculation of the detector geometric acceptance, which defines  the range of the scattered proton momenta accessible to the detectors.
Since the considered physics processes are symmetric in the azimuthal angle, it makes sense to inspect the geometric acceptance as a function of two parameters.
A common choice is the transverse momentum value, $p_T$, and the relative momentum loss of the proton, $\xi = 1-p/p_0$, where $p$ is the scattered proton momentum.

\begin{figure}[htbp]
  \centering
  \includegraphics[width=0.49\linewidth]{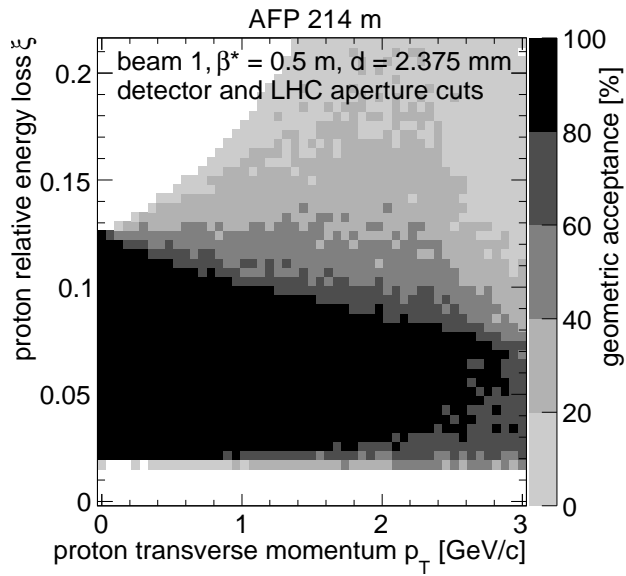}
  \caption{Acceptance of the AFP detectors.}
  \label{fig:acceptance}
\end{figure}

The acceptance of the AFP detectors as a function of the scattered proton $p_T$ and $\xi$  for the nominal settings of the LHC magnets and the beam-detector distance of 15 $\sigma$ is presented in Figure~\ref{fig:acceptance}.
One can see that high acceptance is obtained for $\xi \in [0.02; 0.13]$, \textit{i.e.} high diffractive mass.
To get acceptance for low masses, one needs different LHC optics and vertical detectors.
On the other hand, the AFP detectors provide very good $p_T$ acceptance, allowing the measurements of the full spectrum.

\section{Reconstruction of proton kinematics}

The aim of the forward proton measurement is twofold.
First, one can simply check whether such a proton was present in the event, which would imply the diffractive nature of the process.
Second, measurements of the proton trajectory can be used to determine its momentum, which can be used in the analysis.
In this section we briefly describe the method used for proton kinematics reconstruction with the AFP detectors.

The parameters describing the scattered proton trajectory in the vicinity of the forward proton detector -- the position and elevation angles $(x, y, x', y')$ -- depend on the momentum of the proton emerging from the interaction $(p_x, p_y, p_z)$ as well as on the interaction vertex position $(x_0, y_0, z_0)$.
In the most general case, it would be impossible to reconstruct the momentum of a scattered proton from its trajectory measurement, because the trajectory is described by four parameters, while it depends on six: vertex position and proton momentum.
However, in the case of the nominal LHC optics, the beam spot size is very small.
This makes the vertex dependence sub-leading to the momentum dependence and allows the reconstruction of the momentum.

One of the basic properties of the LHC optics in the neighbourhood of the beam intersection regions is the independence of the horizontal and vertical coordinates.
For example, the horizontal position and elevation angle of a trajectory depend only on the energy of the scattered proton and on the horizontal momentum component $p_x$; they do not depend on the vertical component $p_y$.
One should note that the presence of magnetic moments higher than the quadrupole one in the multipole expansion of the magnetic field introduces such a dependence.
In the case of the LHC magnets the higher order components are a factor of about $10^3$ to $10^4$ smaller than the principal ones \cite{field1,field2,field3} and hence can be neglected.

In the end, the main properties of the optics are of the following form:
\[
(\xi, p_x) \leftrightarrow (x,x') \quad \text{and} \quad (\xi, p_y)
\leftrightarrow (y,y'),
\]
which can be visually presented in terms of the chromaticity plots, see Figures \ref{fig:chromaticity_x} and~\ref{fig:chromaticity_y}.
The grids visible in the plots clearly indicate a one-to-one correspondence between the proton kinematics and its trajectory at the AFP position.
Therefore, it is possible to reconstruct the energy and transverse momentum of a proton from the measurements of its trajectory position and elevation angles.

\twofigures{
  \includegraphics[width=\linewidth]{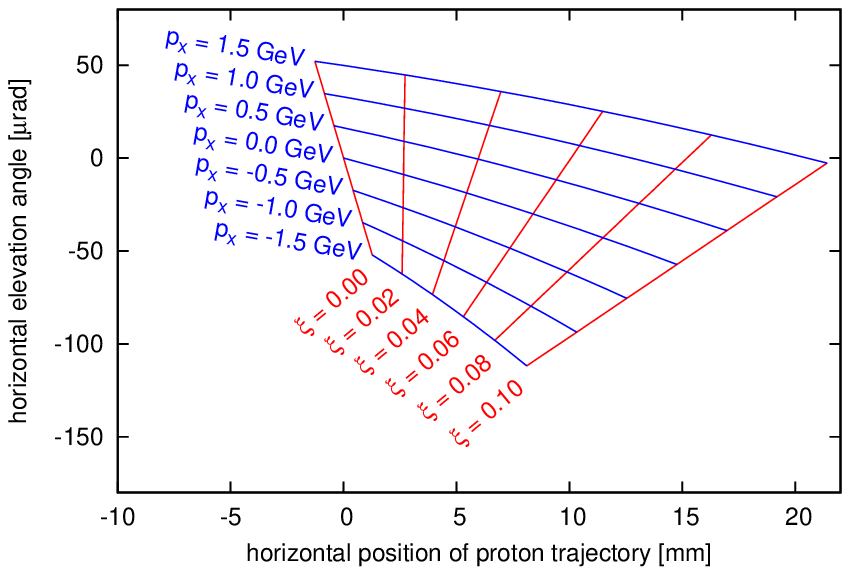}
  \caption{Chromaticity plot for horizontal direction.}
  \label{fig:chromaticity_x}
}
{
  \includegraphics[width=\linewidth]{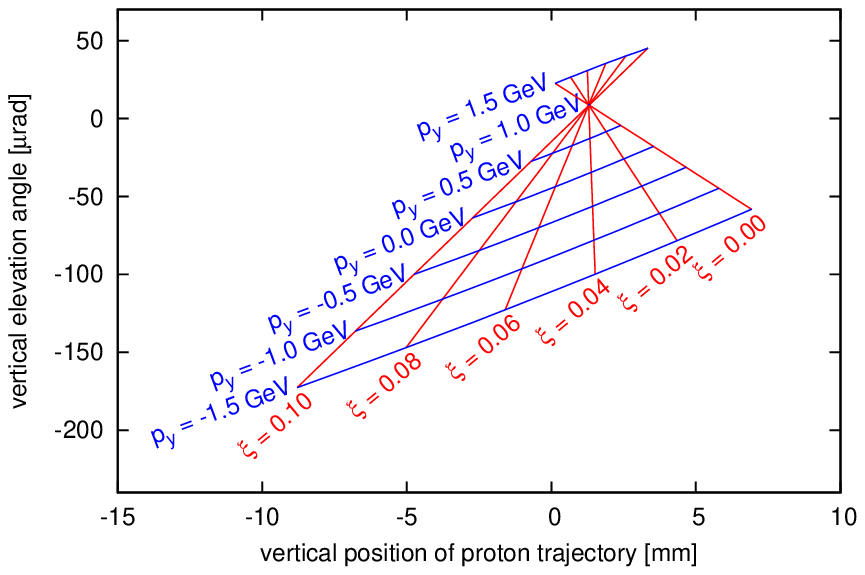}
  \caption{Chromaticity plot for vertical direction.}
  \label{fig:chromaticity_y}
}

For the sake of the alignment analysis in this paper, a simple reconstruction method presented in \cite{Staszewski:2009sw} is used.
For completeness this method is briefly recalled in the following.

The LHC optics between the interaction point and the AFP detectors is parametrised using formulae that assume a linear dependence on the transverse momentum, vertex position and the product of the transverse momentum and the transverse vertex position.
These formulae take into account the independence of the horizontal and vertical directions.
The energy dependence was approximated by polynomials, with ranks chosen in order to provide a negligible error on the transport accuracy with respect to the assumed detector resolution.
For example, the parametrisation formulae for the horizontal position and elevation angle of the proton trajectory at the AFP position are the following:
\[
  x = a(\xi) + b(\xi) \cdot p_x + c(\xi) \cdot x_0 + d(\xi) \cdot p_x \cdot x_0
  + e(\xi) \cdot z_0.
\]
\[
  x' = A(\xi) + B(\xi) \cdot p_x + C(\xi) \cdot x_0 + D(\xi) \cdot p_x \cdot x_0
  + E(\xi) \cdot z_0,
\]

The $\xi$ reconstruction is performed based on the measurement in the horizontal direction only\footnote{Using a more complex method that simultaneously uses the measurements in both directions does not lead to any meaningful improvements.}, by solving the following equation numerically:
\[
  \begin{split}
    \Big(x - a(\xi) - e(\xi) \cdot z_0 - c(\xi) \cdot x_0 \Big)
    \cdot
    \Big( B(\xi) + D(\xi) \cdot z_0 \Big)
    =\\
    \Big(x' - A(\xi) - E(\xi) \cdot z_0 - C(\xi) \cdot x_0 \Big)
    \cdot
    \Big( b(\xi) + d(\xi) \cdot z_0 \Big)
    ,
  \end{split}
\]
which is obtained by extracting $p_x$ from the $x$ and $x'$ parameterisation equations, where $x$ and $x'$ are taken from the measurement in the AFP detectors.

Obviously, in order to obtain a numerical solution, one needs to know the vertex position.
Since usually this in not the case, one should take the centre of the beam spot instead, keeping in mind that such a procedure contributes to the reconstruction resolution.
In this work the stress is placed on the issues related to the apparatus alignment,
while other contributions to the reconstruction error, including the one due to the interaction vertex spread, are neglected (for more details see \cite{Staszewski:2009sw}).
In order to reconstruct the  transverse momentum, the parametrised equations on $x'$ and $y'$ are solved.
It should be pointed out that in principle the equations on $x$ and $y$ could also be used; however, the sensitivity to the transverse momentum is much lower, as is clearly seen in the chromaticity plots.

\section{Effects of misalignment on kinematics reconstruction}

Misalignment of the detectors means that their position is known with a limited precision.
The AFP detector set-up consists of two stations, which results in four degrees of freedom for the position misalignment: $\Delta x_\near$, $\Delta y_\near$, $\Delta x_\far$, $\Delta y_\far$.
Taking linear combinations of these variables, one can define the absolute misalignment, which affects the trajectory position measurement:
\[
  \Delta x = (\Delta x_\near + \Delta x_\far)/2, 
  \quad 
  \Delta y = (\Delta y_\near + \Delta y_\far)/2,
\]
and the relative misalignment, which affects the measurement of the trajectory elevation angles:
\[
  \Delta X = \Delta x_\far - \Delta x_\near, 
  \quad 
  \Delta Y = \Delta y_\far - \Delta y_\near.
\]

Incorrect knowledge about the detectors position will influence the reconstruction procedure.
The numerical results presented in the following part of this paper have been obtained using single diffractive events generated with Pythia \cite{Sjostrand:2007gs, Sjostrand:2006za}.
First, diffractive protons present in these events were transported through the LHC magnetic lattice up to the AFP detector.
Then, their kinematics was reconstructed with appropriate assumptions on the detectors misalignment.
Three cases are considered:
\begin{itemize}
  \item[(a)] absolute horizontal misalignment of 100 $\muup$m ($x$ abs.),
  \item[(b)] relative horizontal misalignment of 100 $\muup$m ($x$ rel.),
  \item[(c)] relative vertical misalignment of 100 $\muup$m ($y$ abs.).
\end{itemize}
The fourth possible case -- absolute vertical misalignment -- has in fact no
impact on the reconstruction. This is due to the reconstruction method, which
does not make use of the vertical trajectory position, but only of the vertical
trajectory angle. The misalignment value has been chosen to 100 $\muup$m, but the effect of any, relatively small, misalignment can be easily estimated, because, to a good approximation, the change in the reconstructed momentum depends linearly on the misalignment value.
It is worth comparing the above benchmark numbers to the alignment precision obtained already at the LHC (for vertical detectors).
ALFA reported \cite{Aad:2014dca} the uncertainty of 1 -- 2 $\muup$m for horizontal position and of about 80 $\muup$m for vertical position.
The TOTEM uncertainties are \cite{Antchev:2013hya}: 5 $\muup$m for horizontal alignment and 30 $\muup$m for vertical alignment.
The big difference between the precision in $x$ and $y$ directions follows from the experimental setup, \textit{i.e.} the geometrical acceptance.
For a detector approaching the beam from below or above, the pattern of elastic events is left-right symmetric, which results in good alignment in the horizontal direction. 

The apparatus misalignment leads to two effects:
\begin{itemize}
  \item[(a)] \textbf{offset} -- the reconstructed values are on average different than the true values,
  \item[(b)] \textbf{spread} -- the reconstructed values are smeared around the average.
\end{itemize}
This requires a short comment.
Naturally, in the presented procedure there is no randomness -- for a given event the reconstructed value will always be the same.
The observed spread of the reconstruction error comes from averaging over all kinematic variables.
For example, different events with the same value of $\xi$ can have different values of $p_x$ and $p_y$.
Then, for each event the reconstruction error on $\xi$ will be different, contributing to the offset and the spread.

Figures~\ref{fig:xi_alignment_offset} and \ref{fig:xi_alignment_spread} present the resulting errors on $\xi$ reconstruction: the offset and the spread, respectively.
One should point out that the misalignment in the vertical direction has no effect, since the $\xi$ reconstruction does not use the vertical proton position.
On the other hand, one can see a contribution due to the horizontal misalignment.
The absolute one leads to an error of 4\%  at small $\xi$, decreasing to about 1\% at high $\xi$ values.
The error due to the relative misalignment has a flatter dependence and slightly varies between -8\% and -6\% within the whole range of the accepted $\xi$ values.
A similar situation is observed for the spread; however, its magnitude is about one order of magnitude smaller.

\twofigures{
  \includegraphics[width=\linewidth]{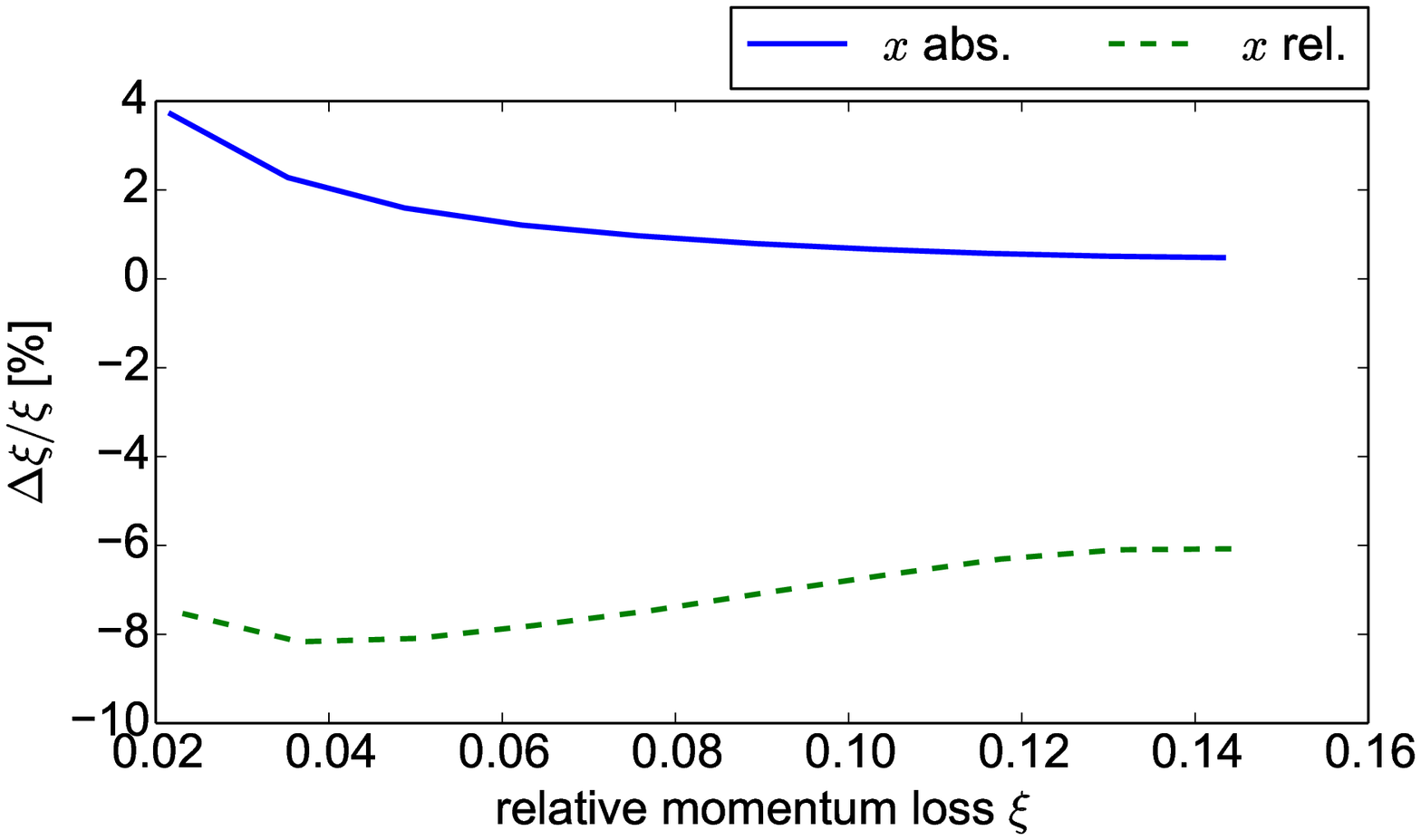}
  \caption{The average reconstruction error of the proton relative momentum loss due to a 100 $\muup$m misalignment.}
  \label{fig:xi_alignment_offset}
}
{
  \includegraphics[width=\linewidth]{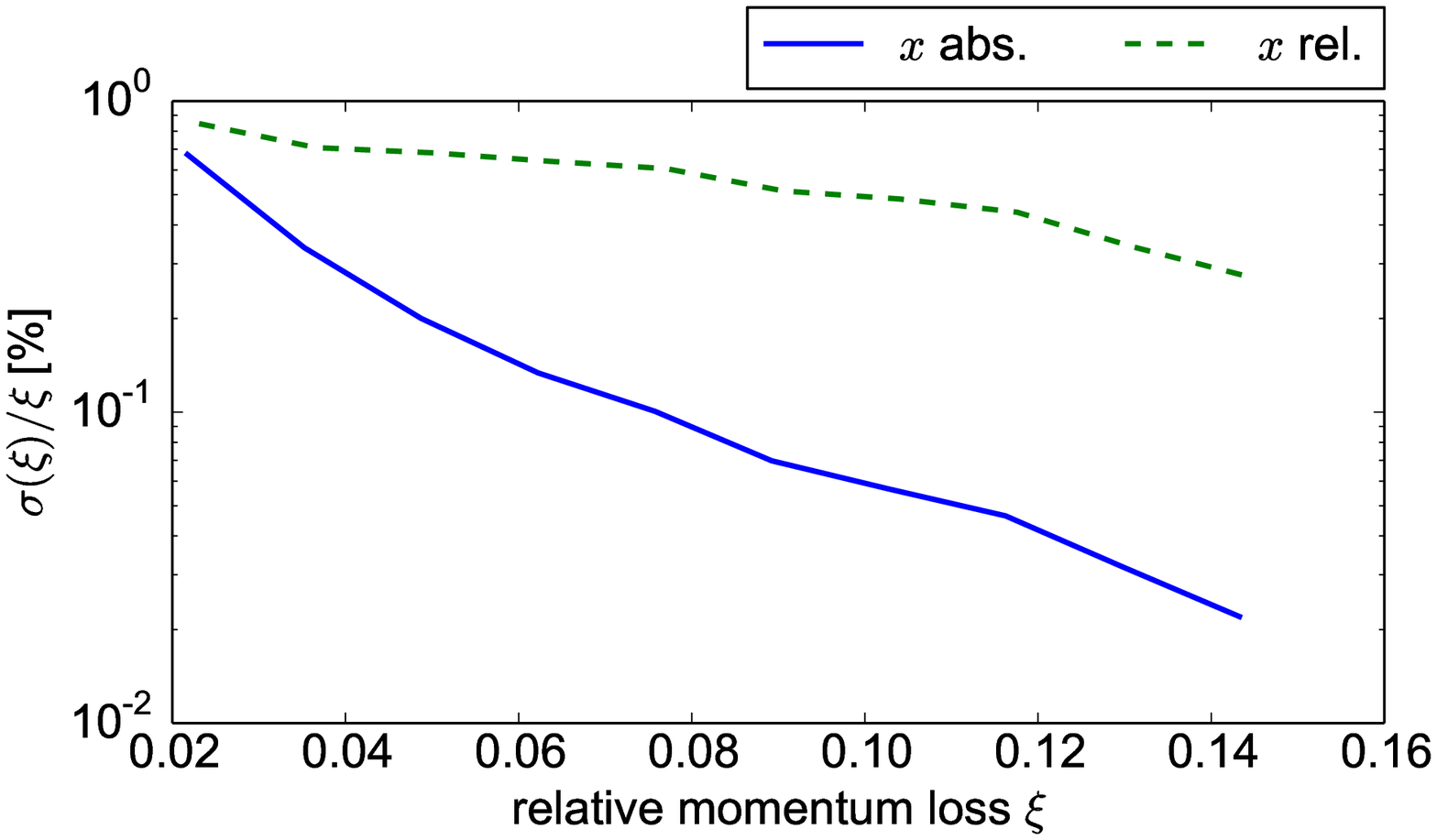}
  \caption{The spread of the relative momentum loss reconstruction error due to a 100 $\muup$m misalignment.}
  \label{fig:xi_alignment_spread}
}

Figures~\ref{fig:px_alignment_offset} and \ref{fig:px_alignment_spread} present the estimated error on the reconstructed $p_x$ value.
The misalignment in the vertical direction does not play a role and is not presented in the plots.
This is also a consequence of the reconstruction method, which does not use vertical components for $p_x$ reconstruction.
The obtained error and spread are dominated by the relative misalignment, with magnitude below 300 MeV and 60 MeV, respectively.

\twofigures{
  \includegraphics[width=\linewidth]{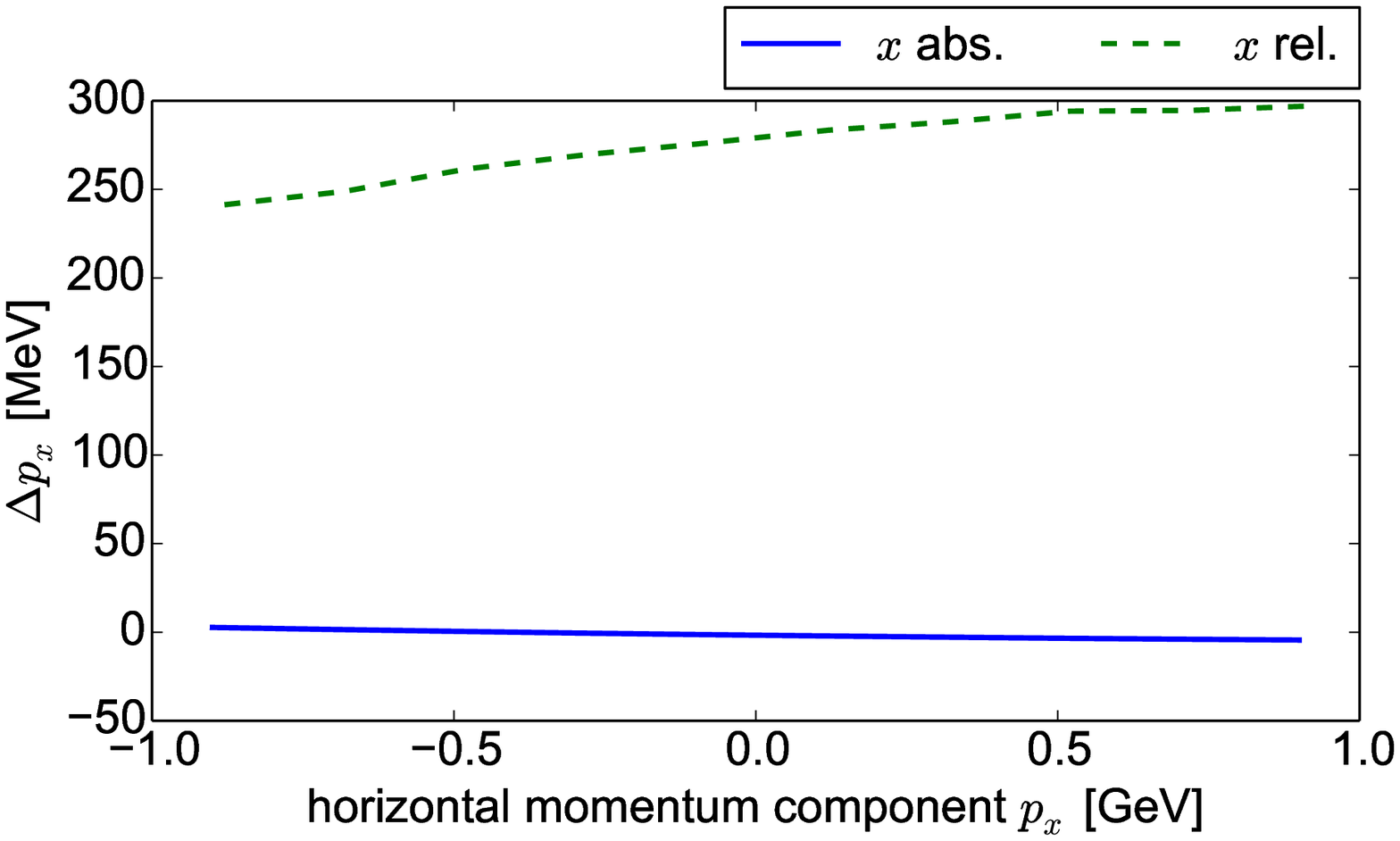}
  \caption{The average reconstruction error of the proton horizontal momentum component due to a 100 $\muup$m misalignment.}
  \label{fig:px_alignment_offset}
}
{
  \includegraphics[width=\linewidth]{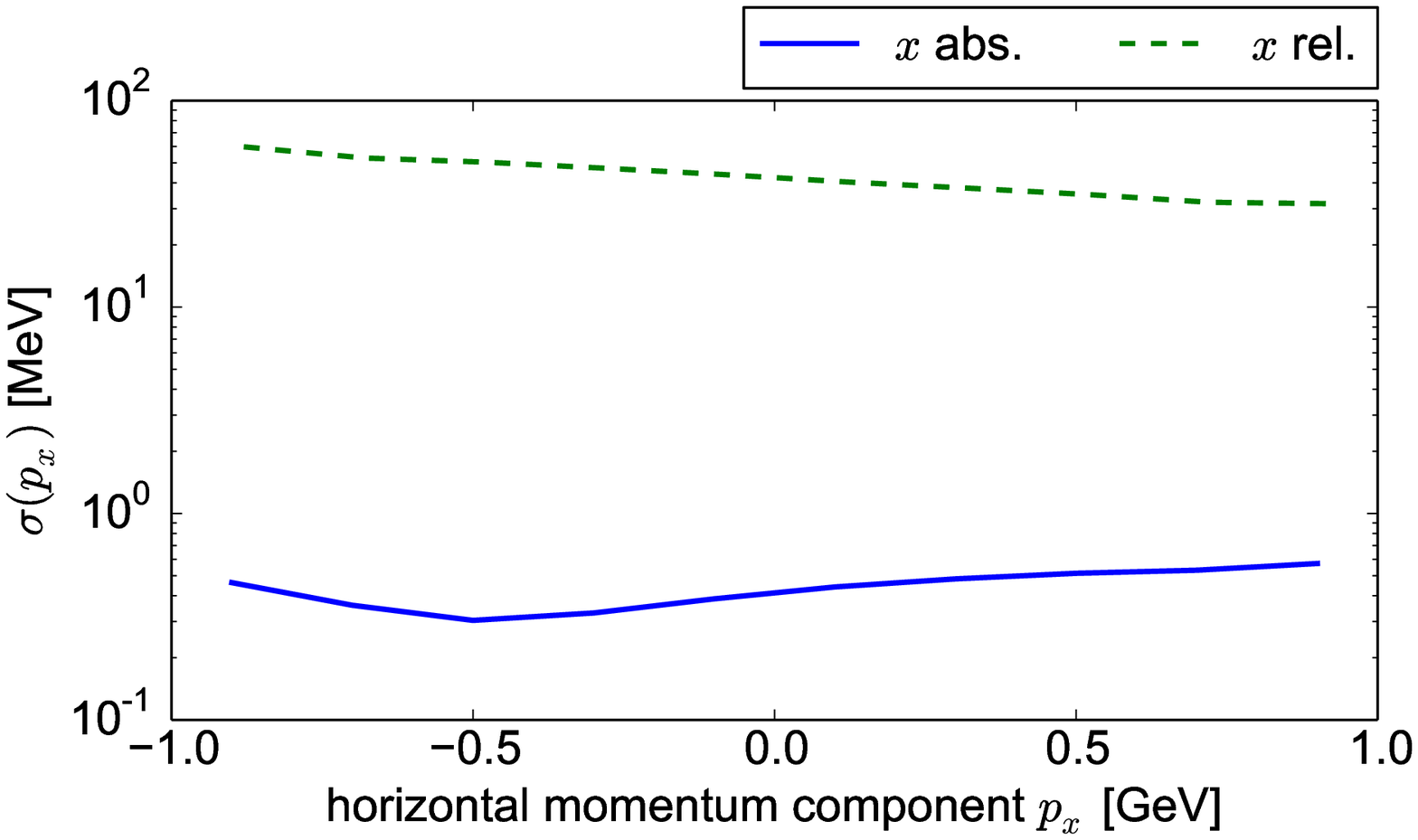}
  \caption{The spread of horizontal momentum reconstruction error due to a 100 $\muup$m misalignment.}
  \label{fig:px_alignment_spread}
}

The results obtained for the vertical momentum reconstruction are presented in Figures~\ref{fig:py_alignment_offset} and \ref{fig:py_alignment_spread}.
Here, the effects of both the horizontal and vertical misalignment are non-negligible.
The horizontal misalignment has a non-zero effect, because it leads to an error on $\xi$, and the reconstructed $\xi$ value is used for the $p_y$ reconstruction.
However, the dominant effect comes from the relative vertical misalignment, which leads to an error of about 330 MeV and spread of about 100 MeV.

\twofigures{
  \includegraphics[width=\linewidth]{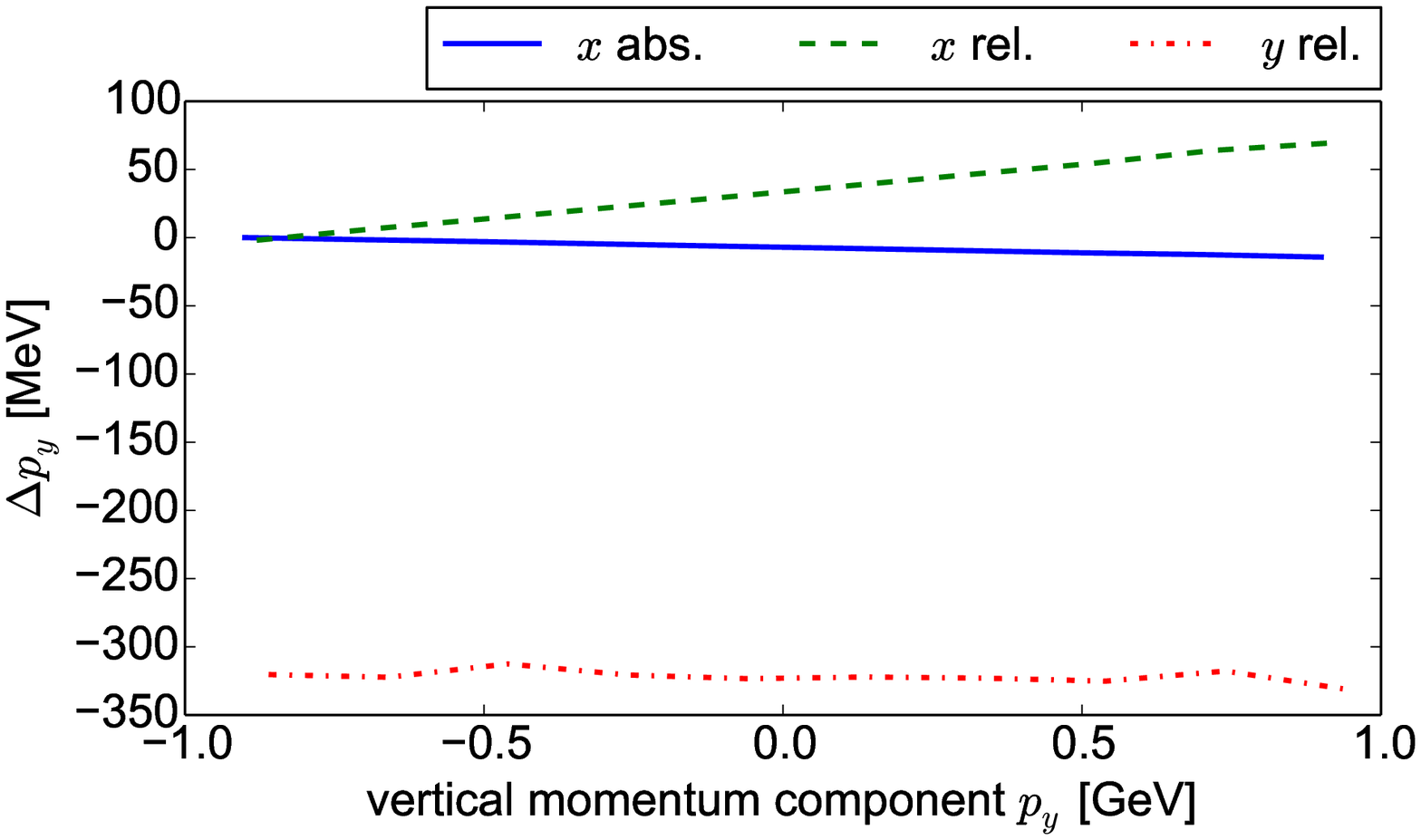}
  \caption{The average reconstruction error of the proton vertical momentum component due to a 100 $\muup$m misalignment.}
  \label{fig:py_alignment_offset}
}
{
  \includegraphics[width=\linewidth]{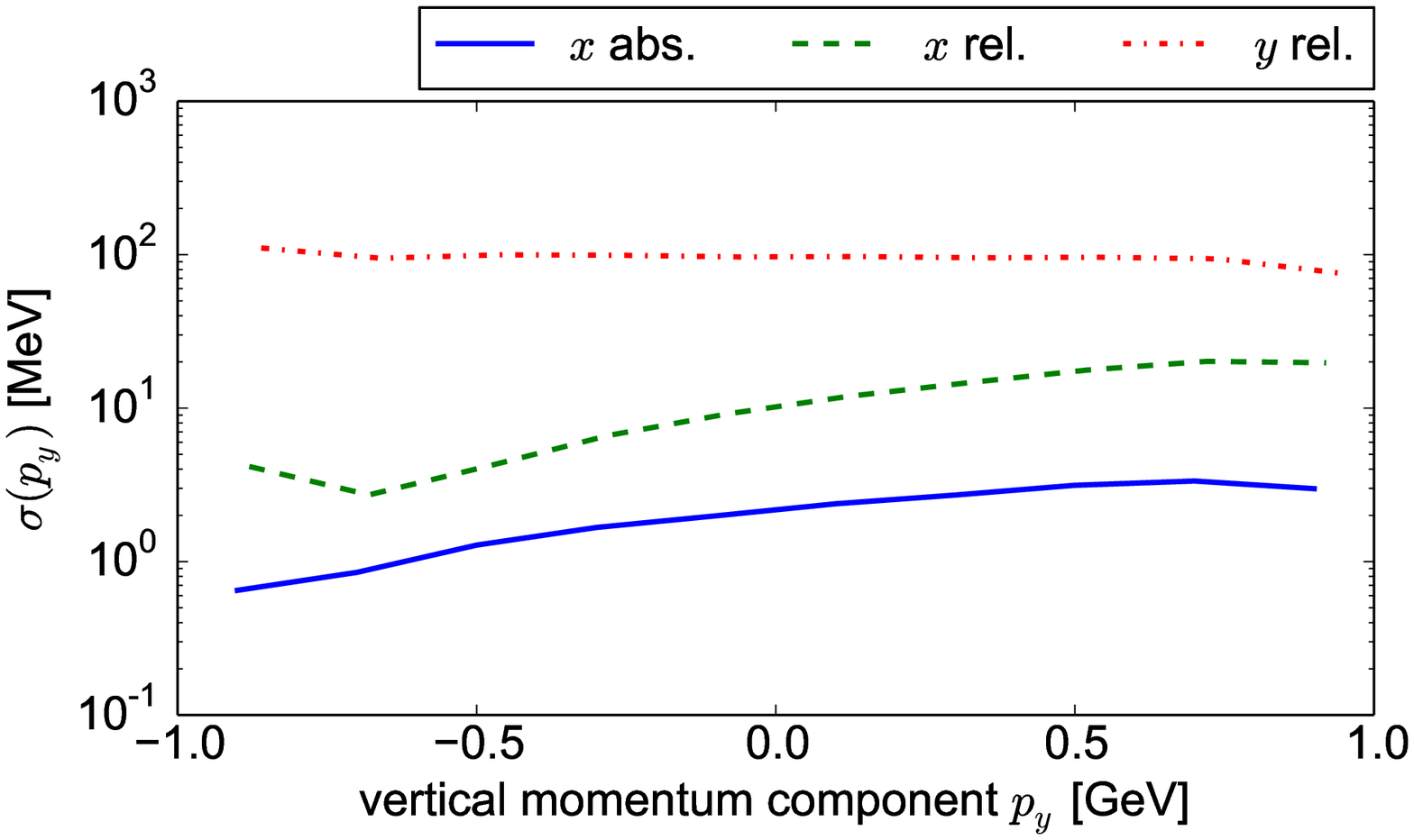}
  \caption{The spread of vertical momentum reconstruction error due to a 100 $\muup$m misalignment.}
  \label{fig:py_alignment_spread}
}

In a physical measurement it is usually not convenient to use directly the horizontal and vertical momentum components, because of the azimuthal symmetry of proton-proton collisions.
Instead, one often uses the four-momentum transfer $t$, related to the magnitude of the transverse momentum ($t\approx p_T^2$).
It is therefore important to study how the reconstruction error would affect this observable.
Obviously, the results cannot contain any additional effect with respect to what has already been presented, but provide a different way of presentation.
The error and spread on the reconstructed four-momentum transfer is presented in Figures~\ref{fig:t_alignment_offset} and~\ref{fig:t_alignment_spread}.
As one could expect, both are dominated by relative misalignments.

\twofigures{
  \includegraphics[width=\linewidth]{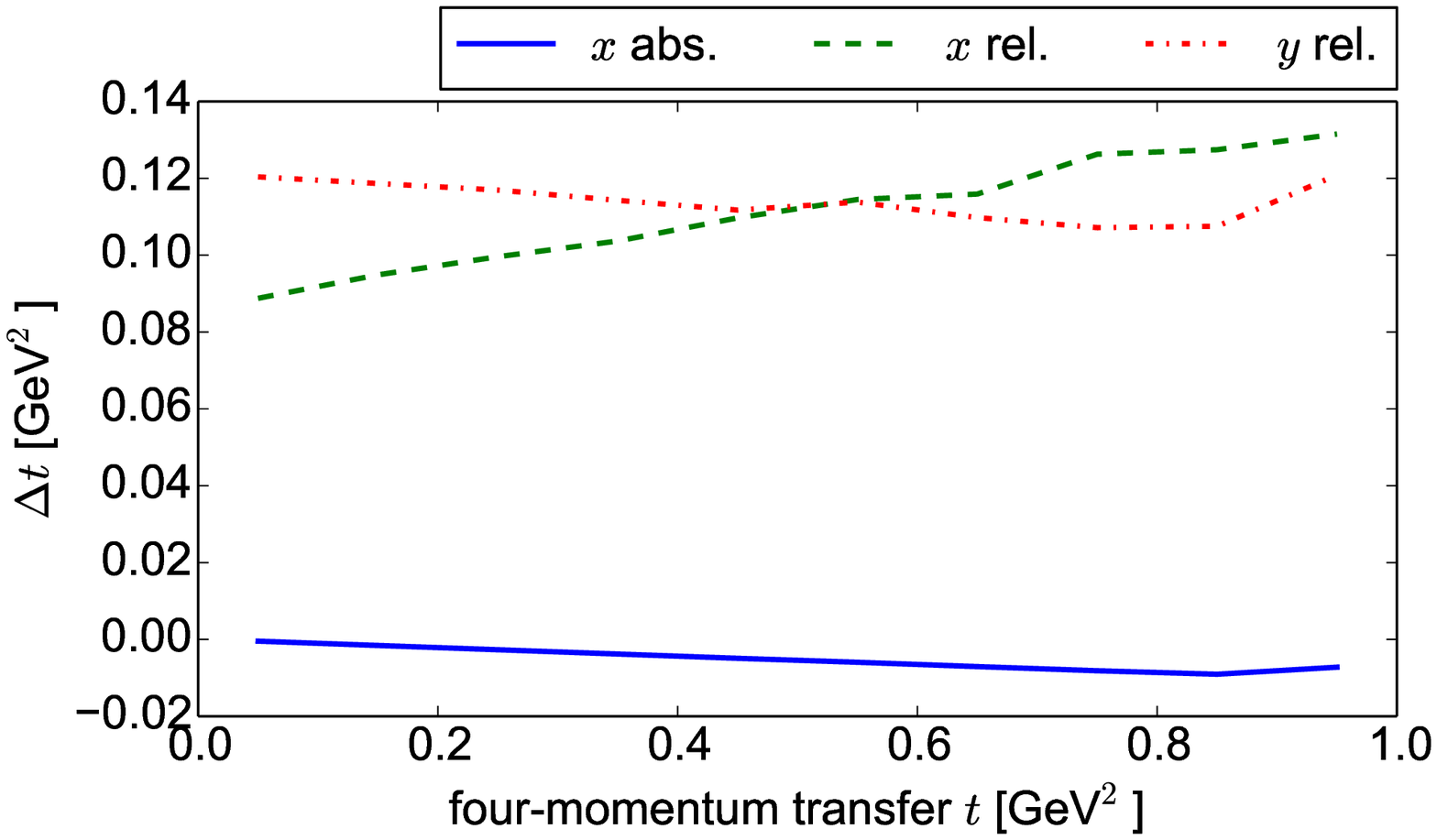}
  \caption{The average reconstruction error of the proton four-momentum transfer due to a 100 $\muup$m misalignment.}
  \label{fig:t_alignment_offset}
}
{
  \includegraphics[width=\linewidth]{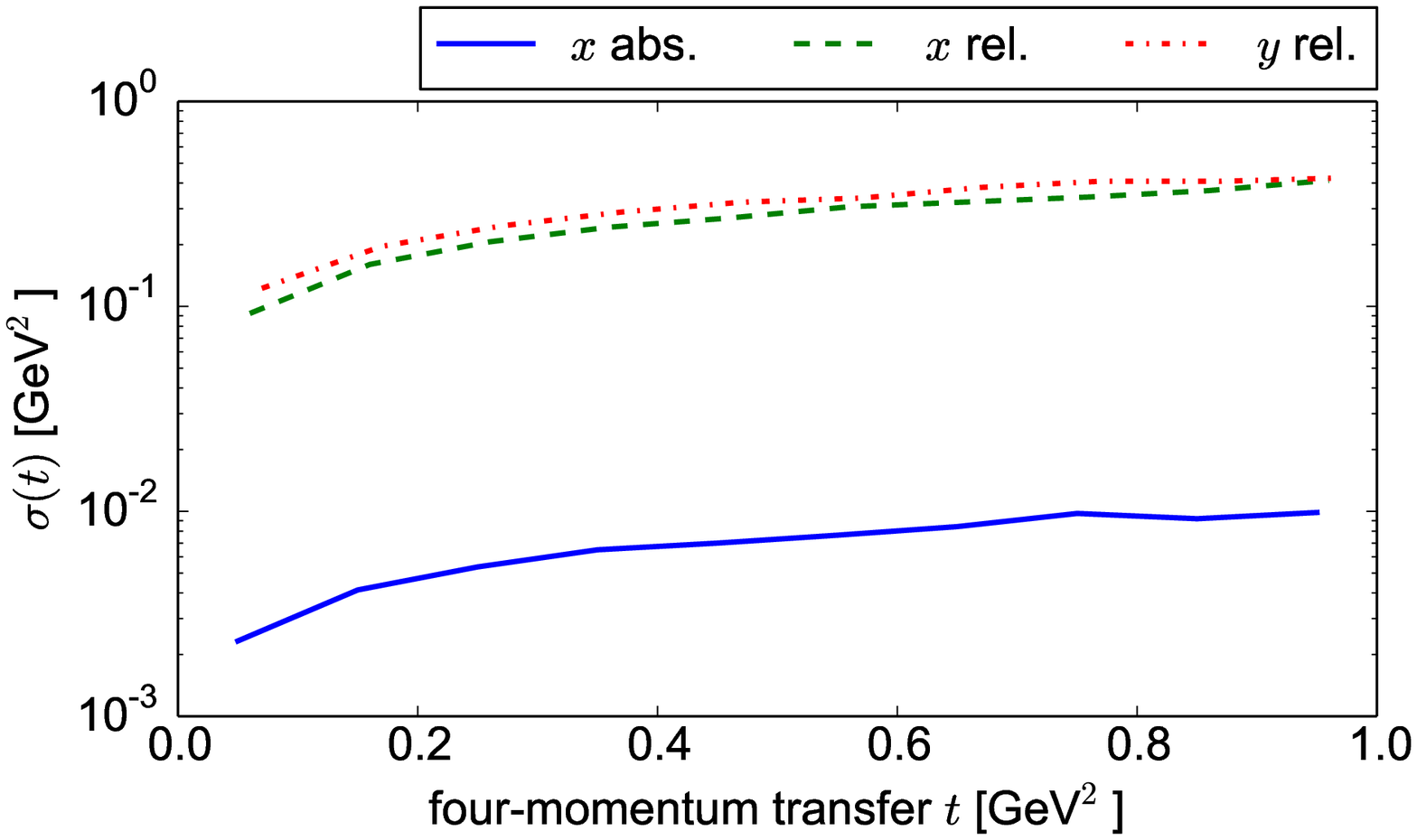}
  \caption{The spread of four-momentum transfer reconstruction error due to a 100
  $\muup$m misalignment.}
  \label{fig:t_alignment_spread}
}

As mentioned before, physics processes are symmetric in the azimuthal angle,
and therefore the measurements of this parameter for processes with a single proton are of no great importance.
However, a meaningful and important observable for events where two forward protons are produced is the azimuthal angle between these two protons.
Therefore, it is relevant to study the misalignment impact on the azimuthal angle at which the proton was scattered.
The results are presented in Figures \ref{fig:phi_alignment_offset} and \ref{fig:phi_alignment_spread}.
Since $\varphi$ is obtained from the reconstructed $p_x$ and $p_y$ values, it it not surprising that its reconstruction is dominated by relative misalignments.
The error and spread obtained for misalignment of 100~$\muup$m are of the order of 1 rad.

\twofigures{
  \includegraphics[width=\linewidth]{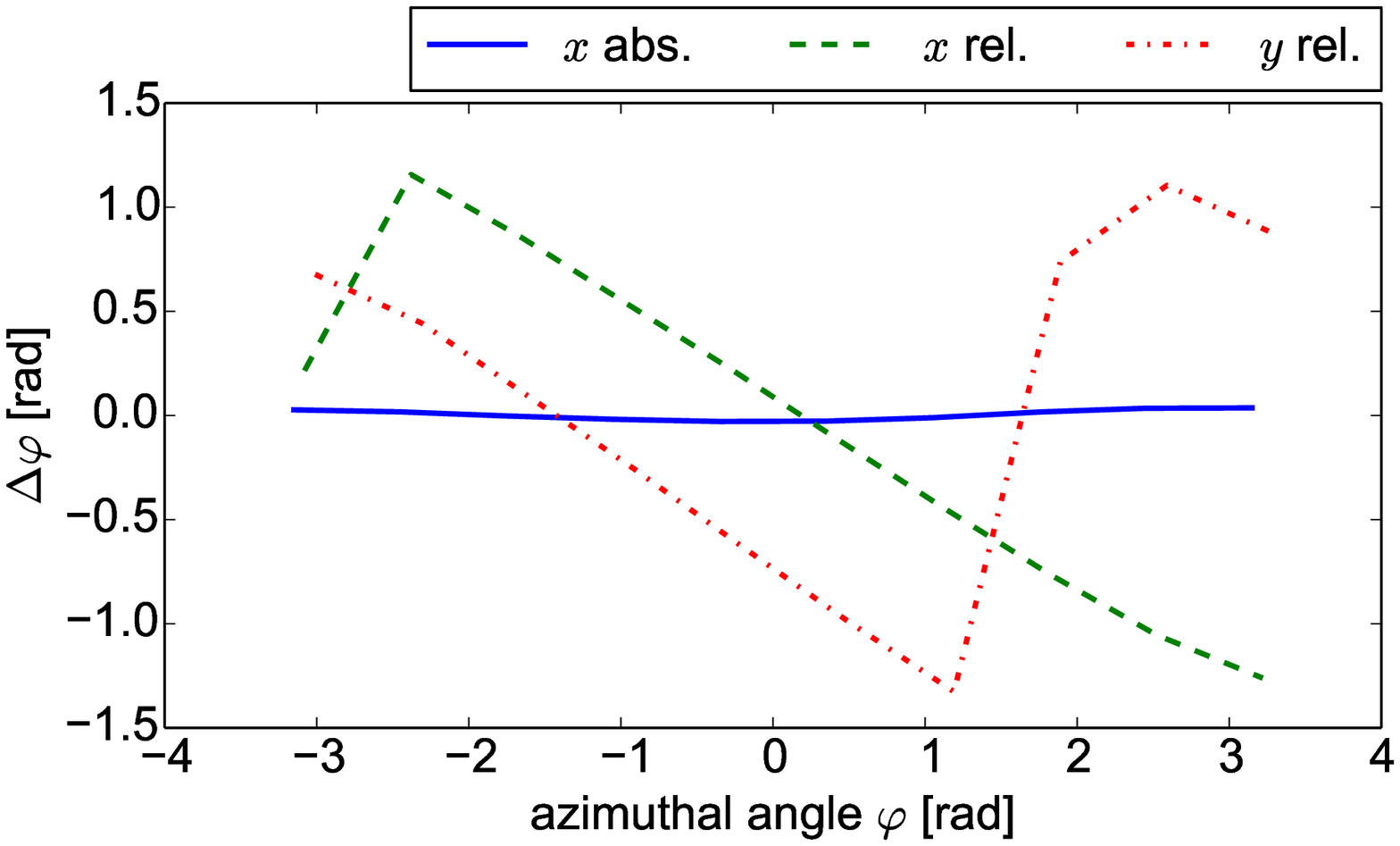}
  \caption{The average reconstruction error of the proton azimuthal angle due to a 100 $\muup$m misalignment.}
  \label{fig:phi_alignment_offset}
}
{
  \includegraphics[width=\linewidth]{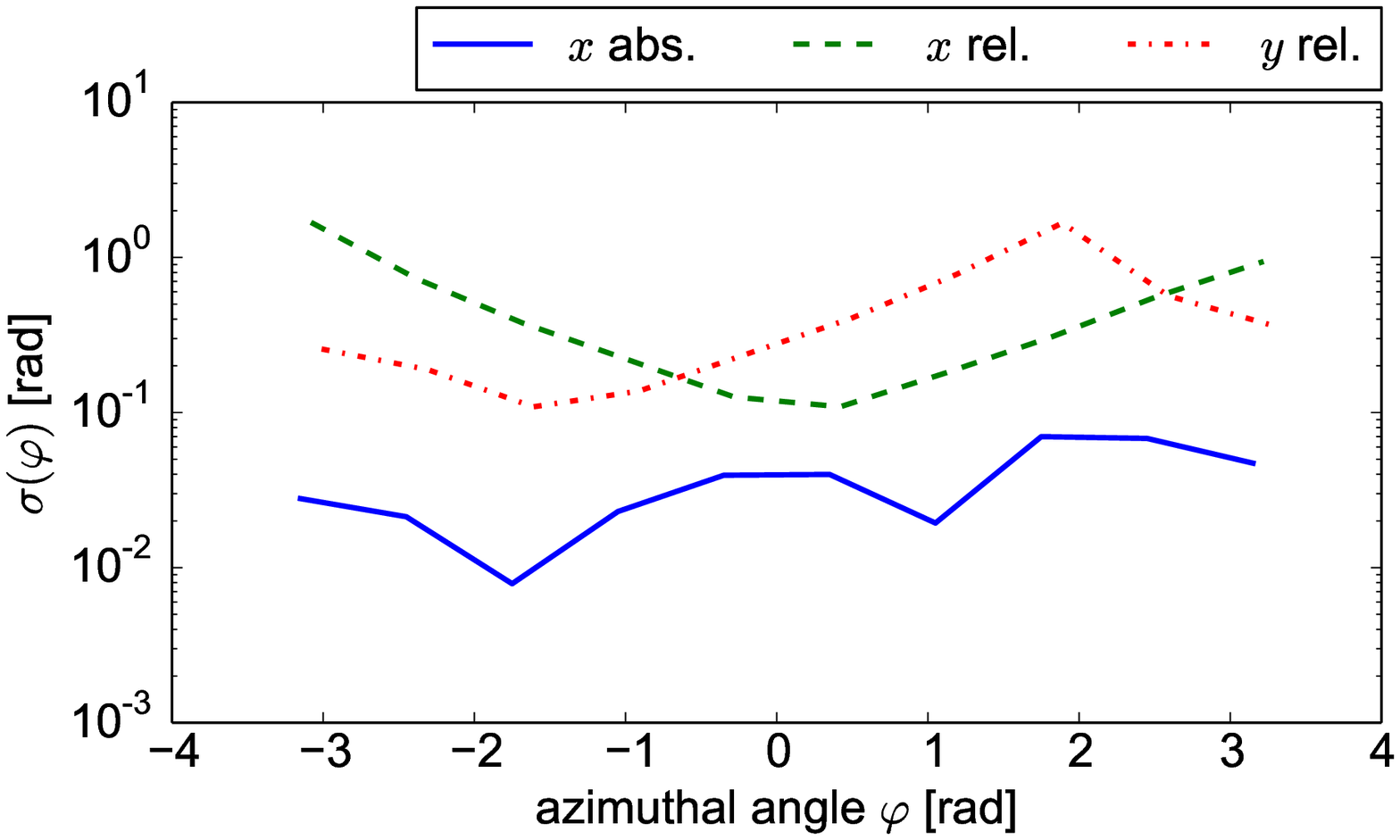}
  \caption{The spread of azimuthal angle reconstruction error due to a 100 $\muup$m misalignment.}
  \label{fig:phi_alignment_spread}
}

The reconstruction of the azimuthal angle naturally depends on the value of the four-momentum transfer.
For small $t$ values it is more difficult to reconstruct $\varphi$.
In particular, for scattering at $t = 0$ the azimuthal angle is indefinite.
Therefore, misalignment is more important for events with smaller $t$ values.
This is illustrated in Figures \ref{fig:phi_t_alignment_offset} and \ref{fig:phi_t_alignment_spread}, where the average error and the error spread on the reconstructed $\varphi$ variable due to relative horizontal misalignment are presented for events in two $t$ ranges: $t < 0.1\ \text{GeV}^2$ and $t > 0.1\ \text{GeV}^2$.
The average error and spread are larger by a factor of about 3 and 2, respectively for $t < 0.1\ \text{GeV}^2$ when compared to $t > 0.1\ \text{GeV}^2$.

\twofigures{
  \includegraphics[width=\linewidth]{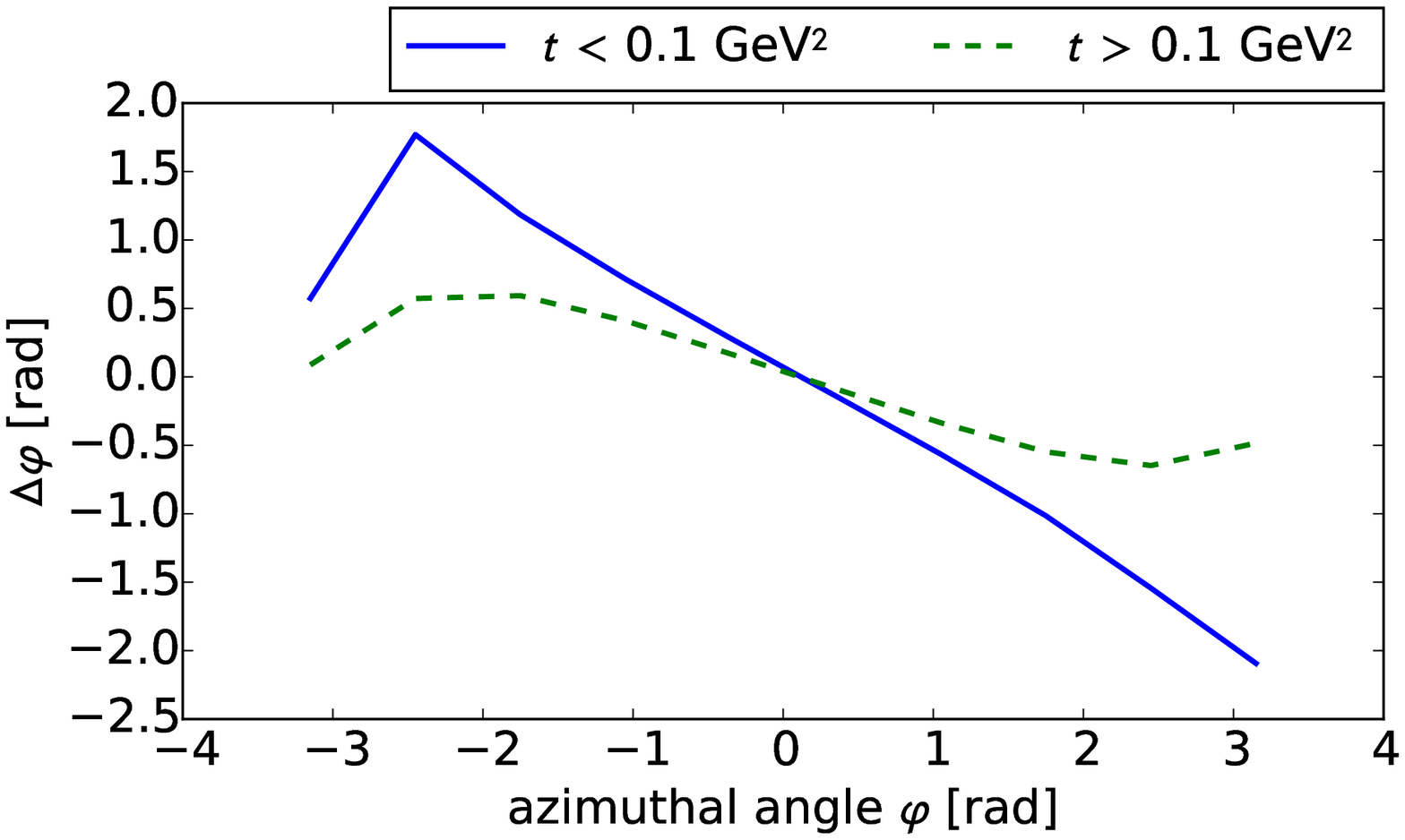}
  \caption{The average reconstruction error of the proton azimuthal angle due to a 100 $\muup$m relative vertical misalignment.}
  \label{fig:phi_t_alignment_offset}
}
{
  \includegraphics[width=\linewidth]{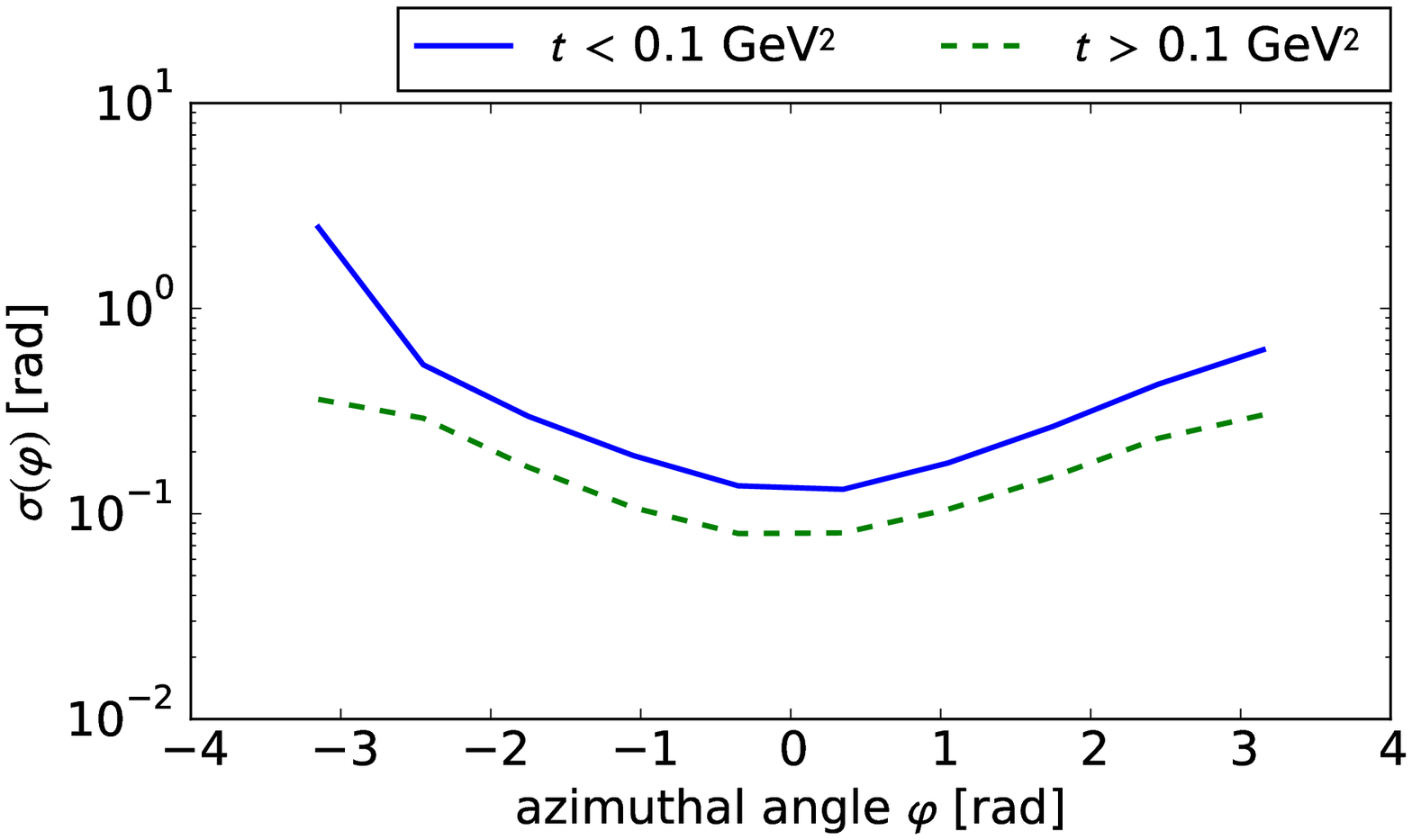}
  \caption{The spread of azimuthal angle reconstruction error due to a 100 $\muup$m relative vertical misalignment.}
  \label{fig:phi_t_alignment_spread}
}

\section{Effects of misalignment on the cross section measurement}

Reconstruction errors are not the only problem caused by the misalignment.
A precise knowledge of the detector position is important even if the detectors are used only for tagging of the scattered protons.
In this case, the misalignment would lead to a wrong estimate of the detector acceptance and in turn to an error on the cross section measurement.

\twofigures{
  \includegraphics[width=\linewidth]{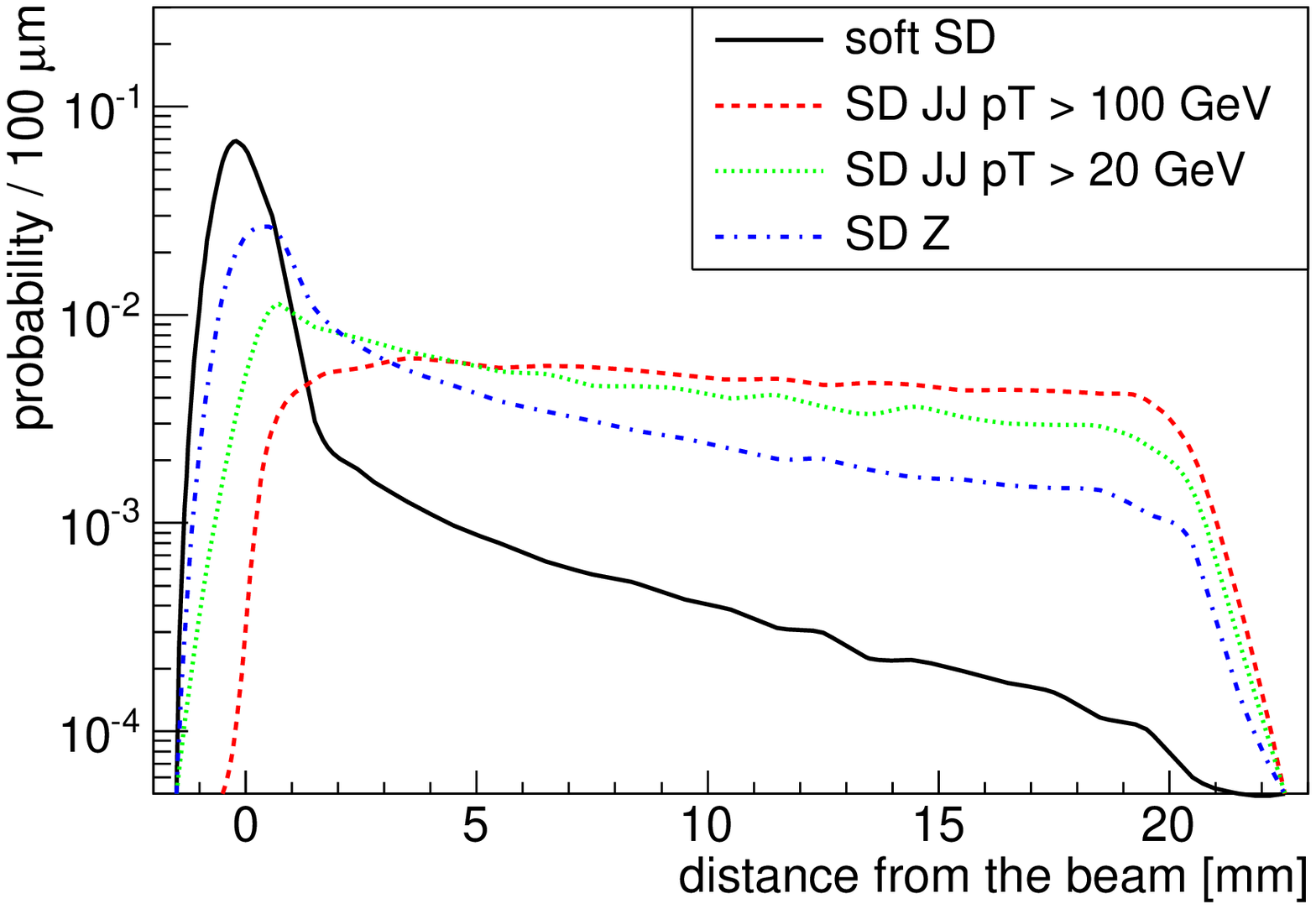}
  \caption{Distribution of the proton's horizontal coordinate at the location of the detectors for various
  diffractive processes.}
  \label{fig:x_distribution}
}
{
  \includegraphics[width=\linewidth]{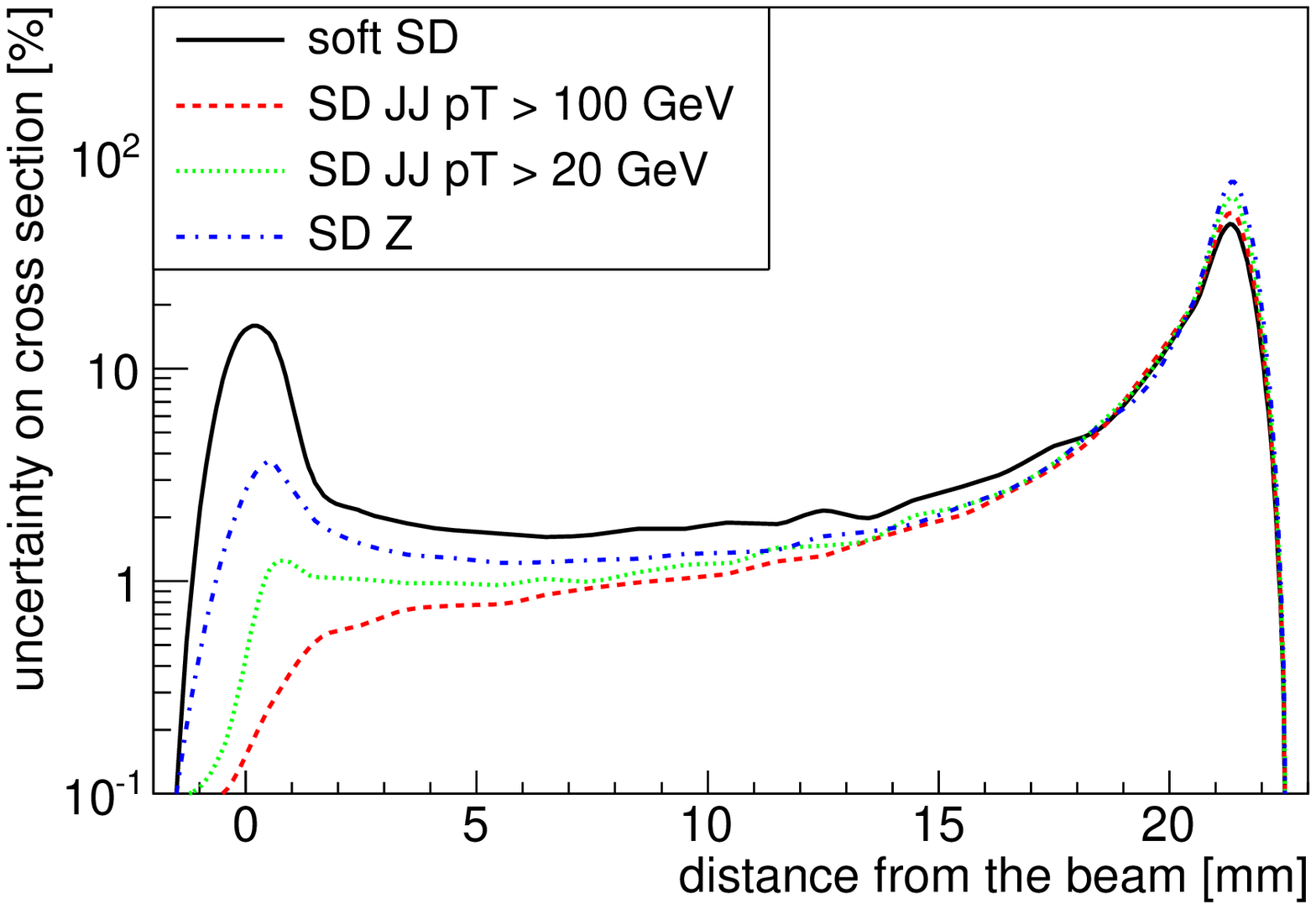}
  \caption{Relative error on the cross section due to a 100 $\muup$m absolute
  horizontal misalignment for several diffractive processes. }
  \label{fig:xsection_error}
}

The vertical misalignment is not relevant for the precise estimation of the detector acceptance, because the acceptance is determined mainly by the distance between the detector and the beam centre\footnote{A negligible, residual effects could be induced by the vertical dimension of the silicon pixel detector.}.
A relevant input to the analysis of this problem is the distribution of the proton horizontal position at the detector location.
This distribution is shown in Figure \ref{fig:x_distribution} for single diffractive dissociation as generated by Pythia and for several hard processes (diffractive jet production with two thresholds on the jet transverse momentum and diffractive production of $Z$ boson) generated with FPMC \cite{Boonekamp:2011ky}.
The peak around zero comes from events with very small momentum loss, for which the proton trajectory position is determined mainly by the transverse momentum acquired in the interaction.
With increasing $x$, the influence of $\xi$ on the position becomes dominant.
The $x$-coordinate distribution flattens off, which reflects the shape of the distribution of the relative momentum loss of diffractive protons, which approximately follows a $1/\xi$ distribution.
Finally, a rapid decrease is observed for $x>20\ \text{mm}$, which is due to the LHC effective aperture.

The error on the cross section measurement caused by the misalignment depends not only on the misalignment value itself but also on the beam-detector distance.
If the detector is very close to the beam, the measurement (tagging) is performed in the steeply falling region of the distribution.
Therefore, even a small change in the distance would have a large effect.
When the detector performs the measurement in the flat region of the distribution, the sensitivity to misalignment gets much smaller.
Towards the boundary of the acceptance region the error increases again, because the observed cross section is already quite small and even a tiny change is significant.
These considerations are summarised in Figure~\ref{fig:xsection_error}, which presents the error obtained for a misalignment of 100 $\muup$m as a function of the detector distance from the beam centre.

\section{Rotations}

Besides the linear misalignment discussed above, it is also possible that an unknown rotation is introduced to the detector.
In principle, one could consider rotations in three planes: $(x,y)$, $(x,z)$ and $(y,z)$.
However, in the case of rotations by a small angle $\theta$ in the $(x,z)$ and $(y,z)$ plane, the $x$ and $y$ trajectory positions are only modified by a term of $\mathcal{O}(\theta^2)$, which can be safely neglected.
The change in the $z$ position is also negligible, since it is of the order of $\theta$ times the size of the sensor (about 20 mm), which should be compared to the 8 m distance between two AFP stations.
Therefore, in the following we focus only on rotations in $(x,y)$ plane.

Obviously, the rotation can be of different mechanical origin.
It may be related to the design of the detector station -- the presence of an unwanted inclination in the position of the Roman pot, or wrong mechanical support.
However, one should keep in mind that a rotation around a given axis is equivalent to a rotation around a different, parallel axis and a translation.
A rotation around an axis far from the detector sensor will give a large contribution to the offset of the active area.
Since the translational misalignment has already been discussed, it is sufficient to consider rotations w.r.t. an axis close to the detector.
The choice of this axis would affect the reconstruction -- for protons close to the rotation axis the reconstruction error should be smaller, and it will increase with distance from the axis.

Contrary to the translational misalignment, it is not possible to define absolute and relative rotations.
One can only apply the rotation to each of the stations.
In the following, four cases are considered:
\begin{itemize} 
  \item[(a)] rotation of the near station around the centre of the diffractive pattern at the edge close to the beam, 
  \item[(b)] rotation of the far station around the centre of the diffractive pattern at the edge close to the beam, 
  \item[(c)] simultaneous rotation of both stations around the centre of the diffractive pattern at the edge close to the beam,
  \item[(d)] rotation of the near station around the centre of the beam.
\end{itemize}
Like in the previous section, the impact on the reconstruction of different kinematic variables will be presented.
The magnitude of all rotations is set to 1 milliradian.
This can be compared to the uncertainties reported by ALFA \cite{Aad:2014dca} and TOTEM \cite{Antchev:2013hya}: 0.1 and 0.5 mrad, respectively.
Figures \ref{fig:xi_rotation_offset} and  \ref{fig:xi_rotation_spread} present the offset and the spread on the reconstructed $\xi$ value, respectively.
Rotation of the near station has a very similar effect to that induced by the rotation of the far station, except that the offsets have opposite signs.
This is a reminiscence of the fact that the change of the trajectory elevation angles has a greater impact on the $\xi$ reconstruction than the position change.
The change of the trajectory elevation angles due to the rotation of the far station has an opposite sign to the change due to the rotation of the near station.
A similar argument is appropriate also for the spread.
It is interesting to notice that the errors in the case when both stations are rotated are much smaller than when only a single station is rotated.
The difference between the different origins of rotation is non-negligible only for the offset -- as expected the rotation around the beam centre results in larger errors.
The magnitude of the error is below 1 \% for the offset and below 1 $\permil$ for the spread.

\twofigures{
  \includegraphics[width=\linewidth]{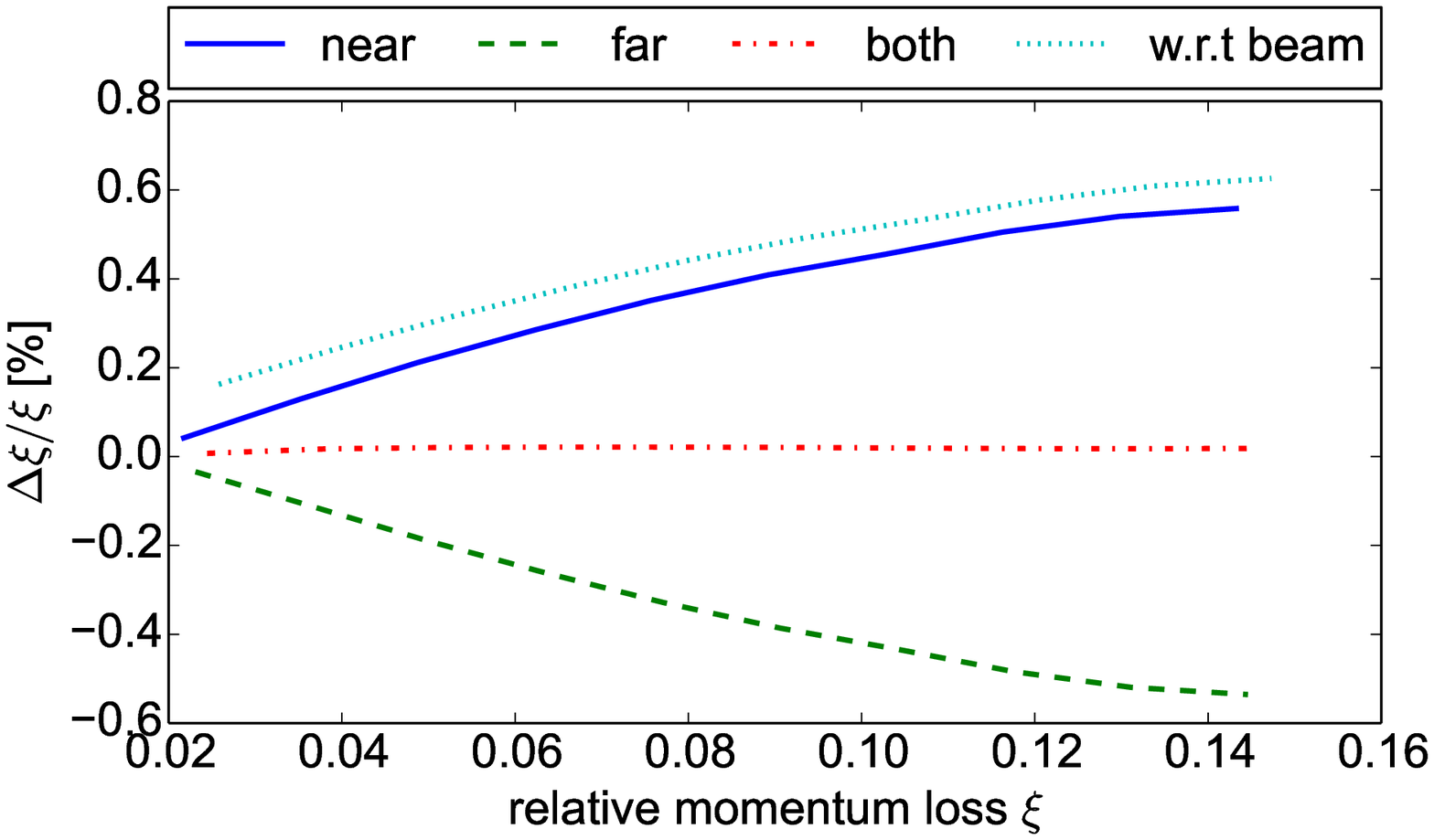}
  \caption{The average reconstruction error of the proton relative momentum loss due to a 1 mrad rotation of the detectors.}
  \label{fig:xi_rotation_offset}
}
{
  \includegraphics[width=\linewidth]{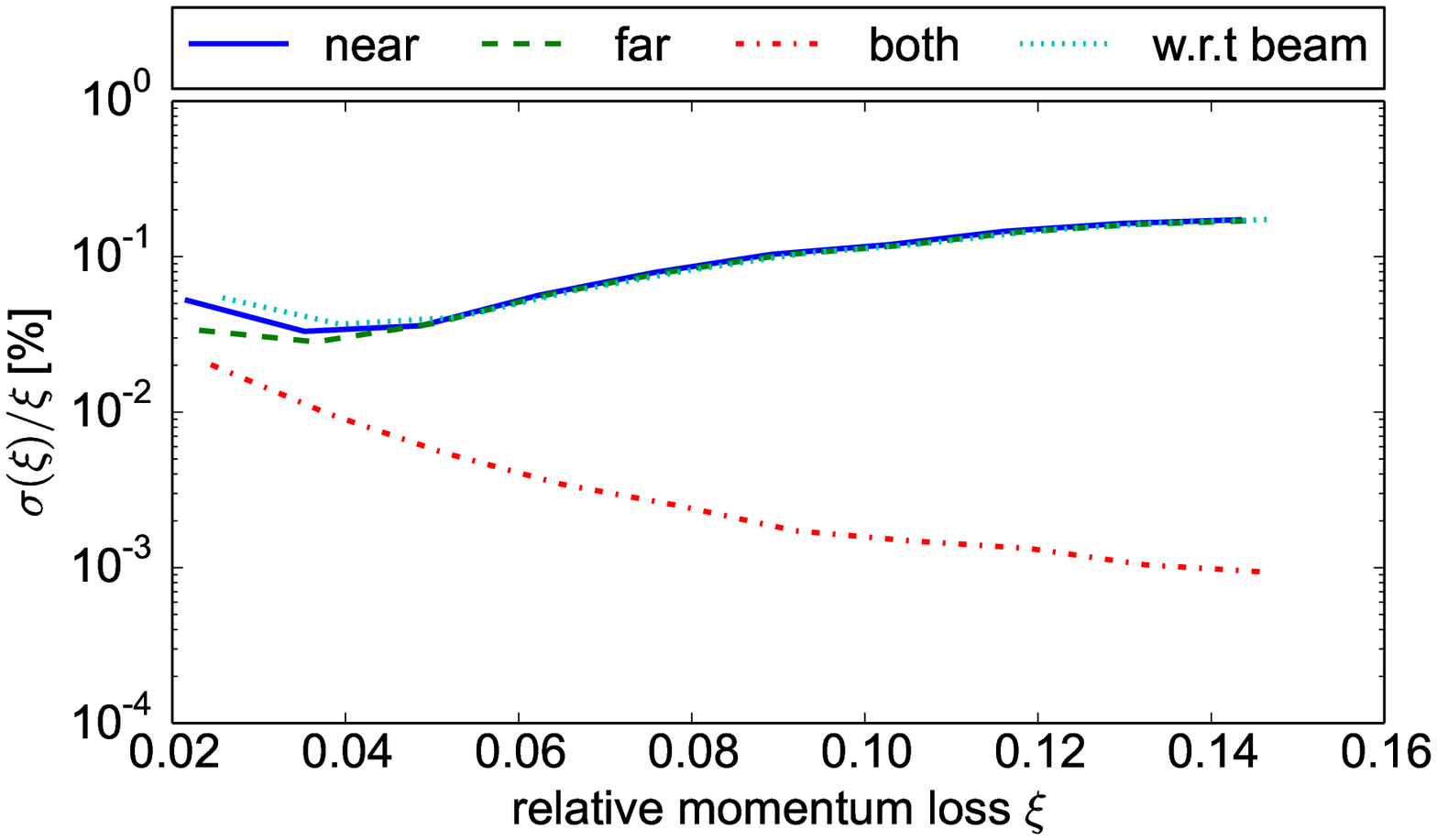}
  \caption{The spread of the reconstruction error of the proton relative momentum loss due to a 1 mrad rotation of the detectors.}
  \label{fig:xi_rotation_spread}
}

Figures \ref{fig:px_rotation_offset} and \ref{fig:px_rotation_spread} present the errors in the case of the $p_x$ reconstruction.
Both the offset and the spread are of the order of few MeV, when one station is rotated.
The rotation around a more distant point increases slightly the offset.
The simultaneous rotation of both stations leads to negligible errors.
The case for the reconstruction of the vertical momentum component, presented in Figures \ref{fig:py_rotation_offset} and \ref{fig:py_rotation_spread}, is very similar; however, the errors are larger by approximately a factor of two.

\twofigures{
  \includegraphics[width=\linewidth]{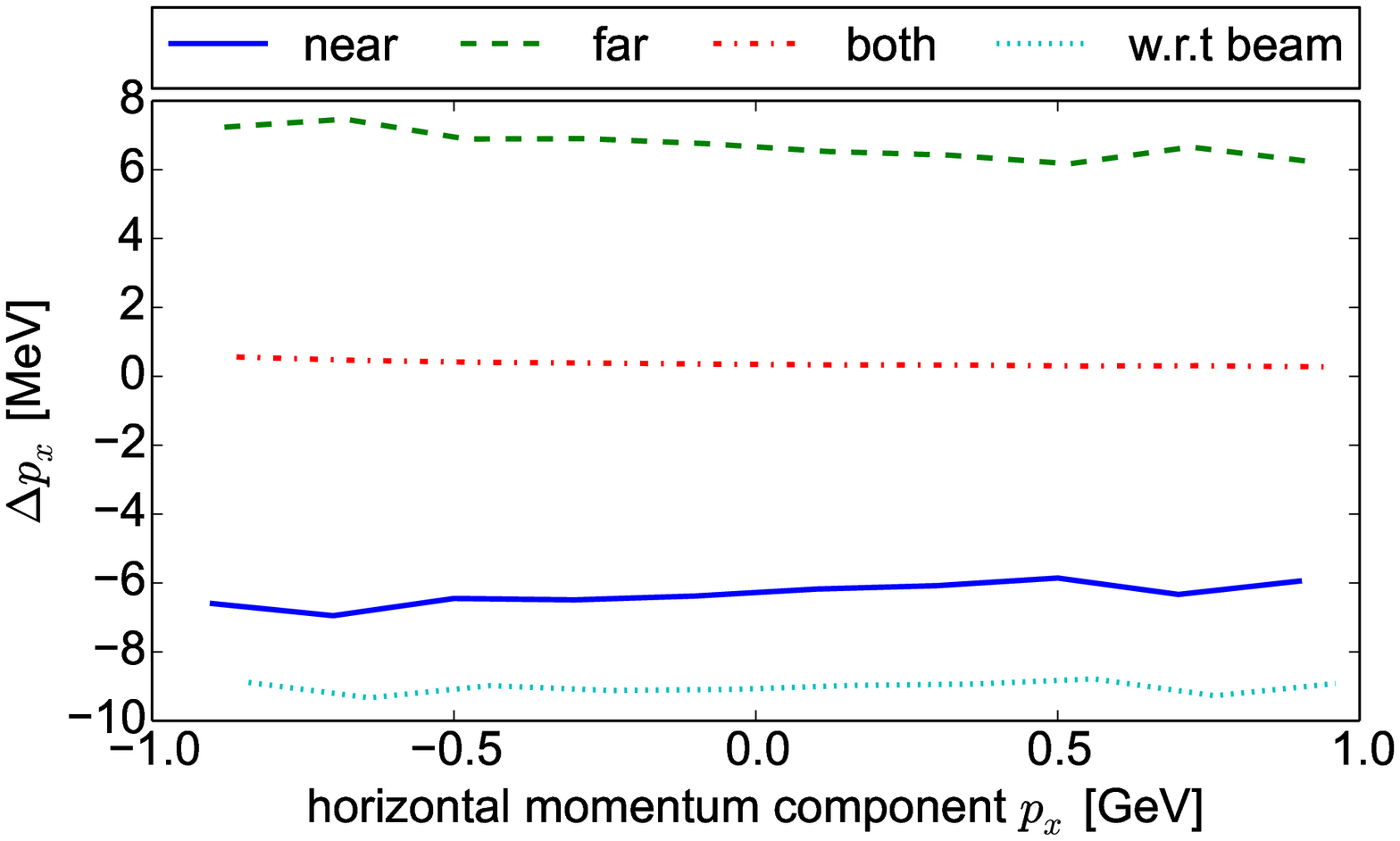}
  \caption{The average reconstruction error of the horizontal component of the proton momentum due to a 1 mrad rotation of the detectors.}
  \label{fig:px_rotation_offset}
}
{
  \includegraphics[width=\linewidth]{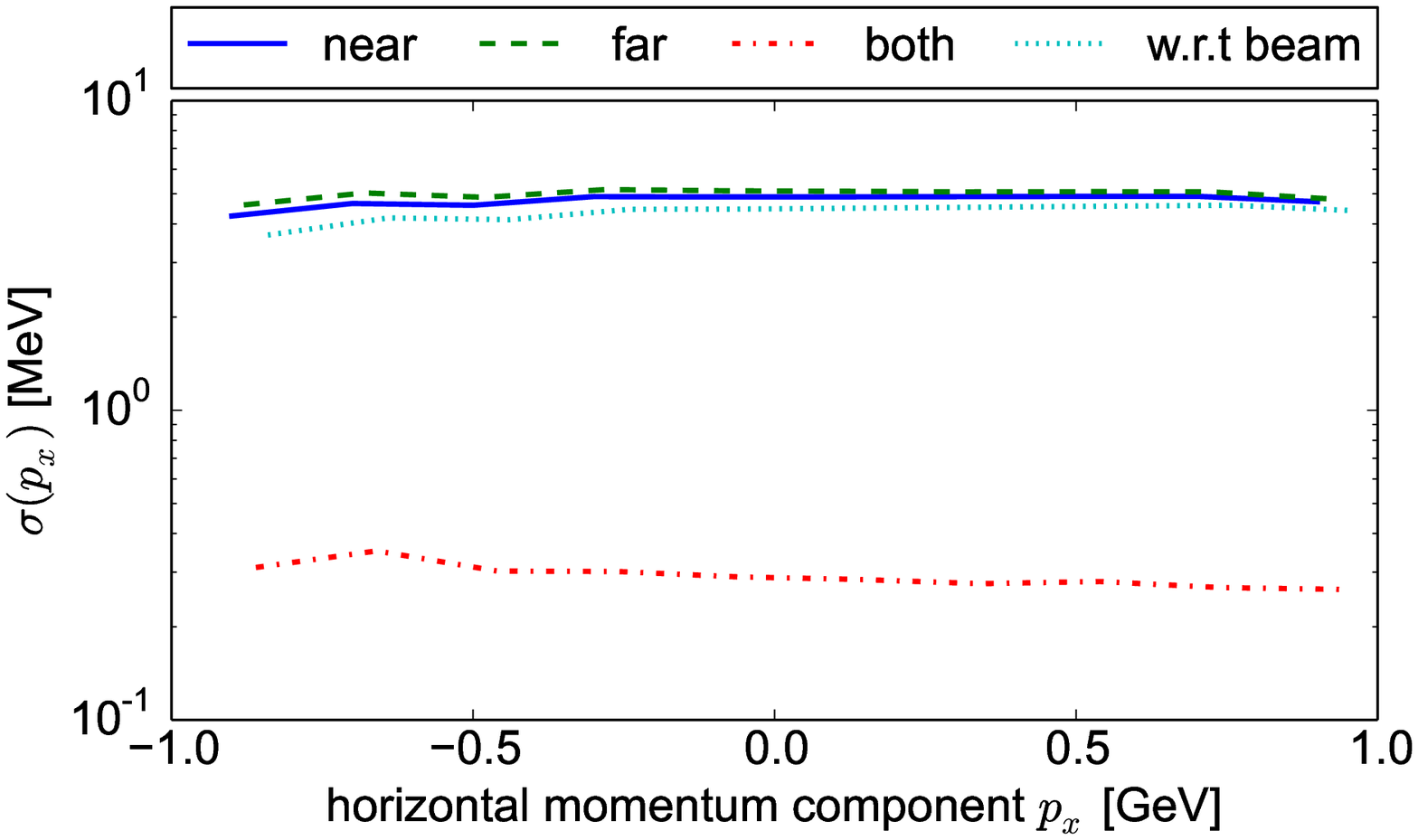}
  \caption{The spread of the reconstruction error of the horizontal component of the proton momentum due to a 1 mrad rotation of the detectors.}
  \label{fig:px_rotation_spread}
}

\twofigures{
  \includegraphics[width=\linewidth]{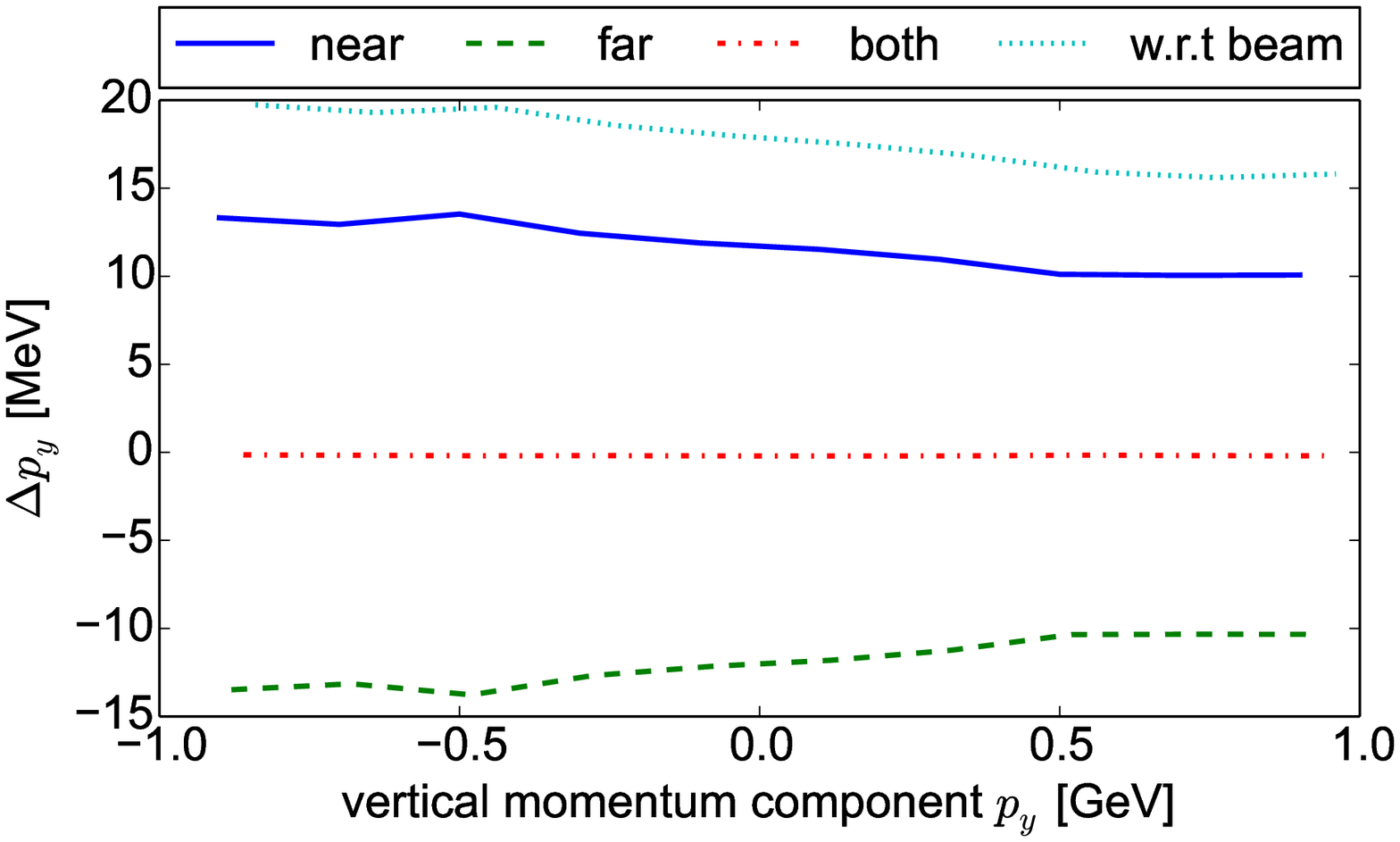}
  \caption{The average reconstruction error of the vertical component of the proton momentum due to a 1 mrad rotation of the detectors.}
  \label{fig:py_rotation_offset}
}
{
  \includegraphics[width=\linewidth]{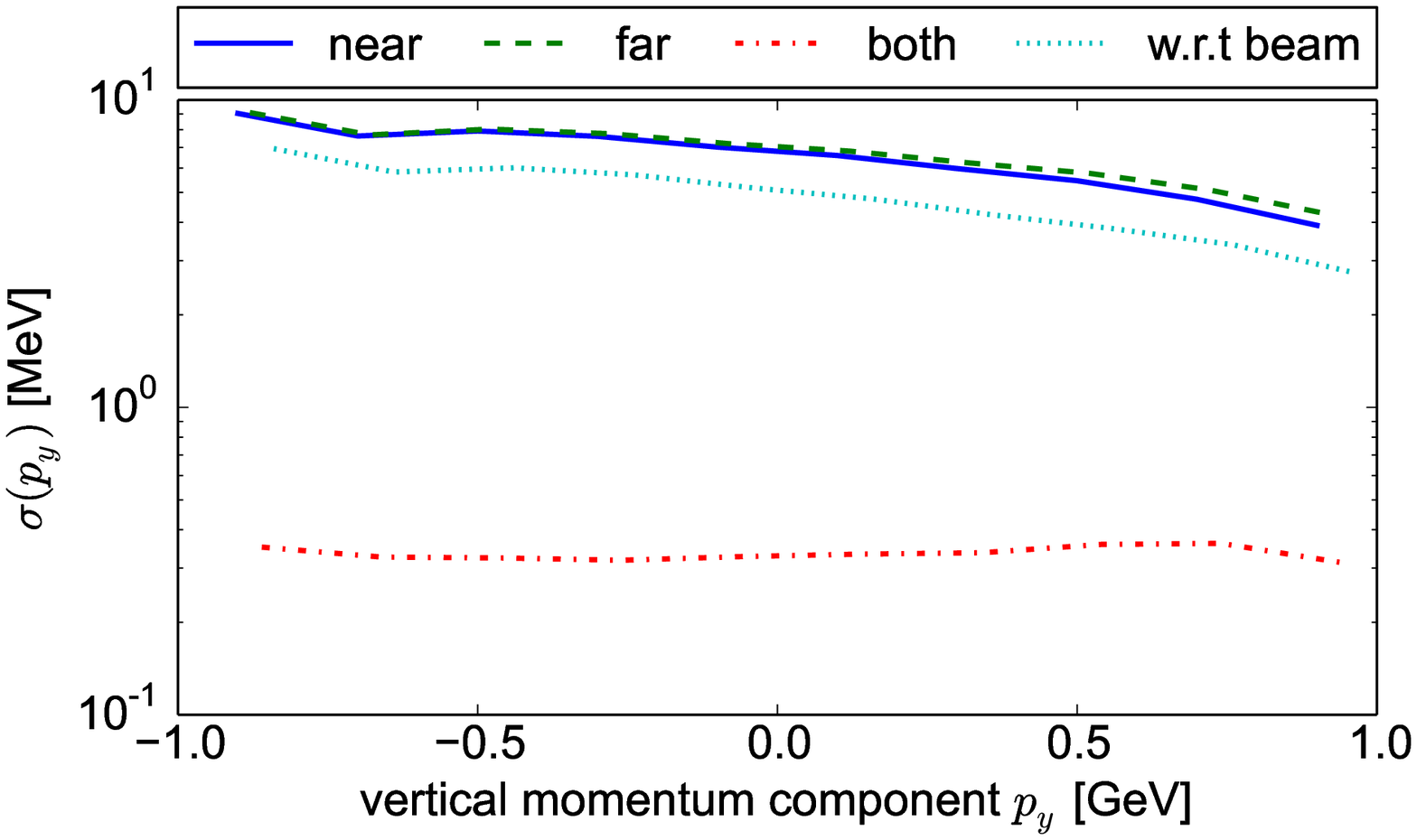}
  \caption{The spread of the reconstruction error of the vertical component of the proton momentum due to a 1 mrad rotation of the detectors.}
  \label{fig:py_rotation_spread}
}

Figures \ref{fig:t_rotation_offset} and \ref{fig:t_rotation_spread} present the error in the case of the four-momentum transfer reconstruction.
Also here, the simultaneous rotation of both stations results in negligible errors.
The rotation of a single station leads to an offset of the order of $10^{-3}$ GeV$^{2}$ and a spread of the order of $10^{-2}$ GeV$^{2}$.
The offset is smaller than the spread due to the fact that a given value of $t$ can be obtained for different combinations of $p_x$ and $p_y$ values.
The individual offsets on $p_x$ and $p_y$ nearly cancel on the average.

\twofigures{
  \includegraphics[width=\linewidth]{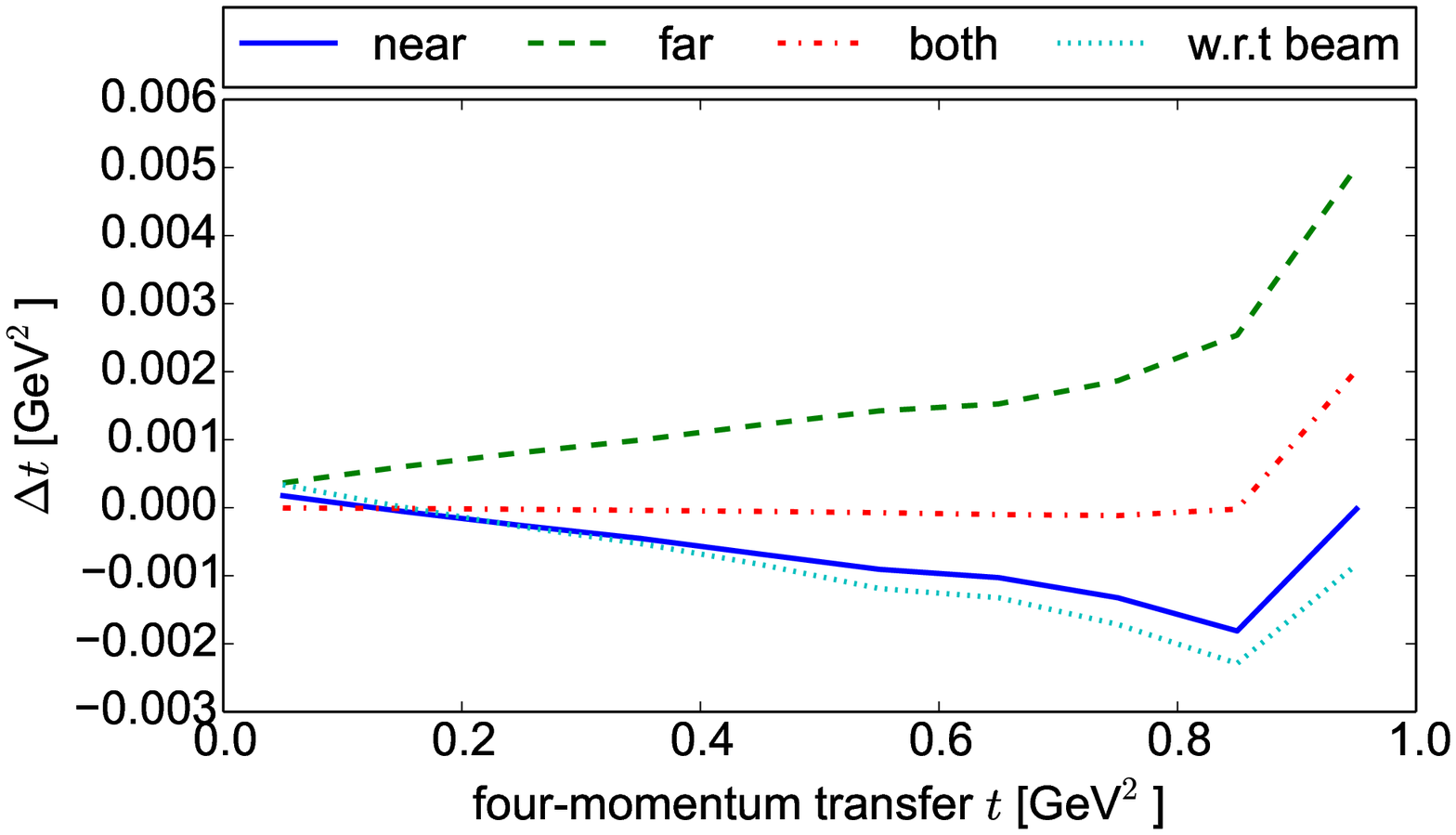}
  \caption{The average reconstruction error of the four-momentum transfer due to a 1 mrad rotation of the detectors.}
  \label{fig:t_rotation_offset}
}
{
  \includegraphics[width=\linewidth]{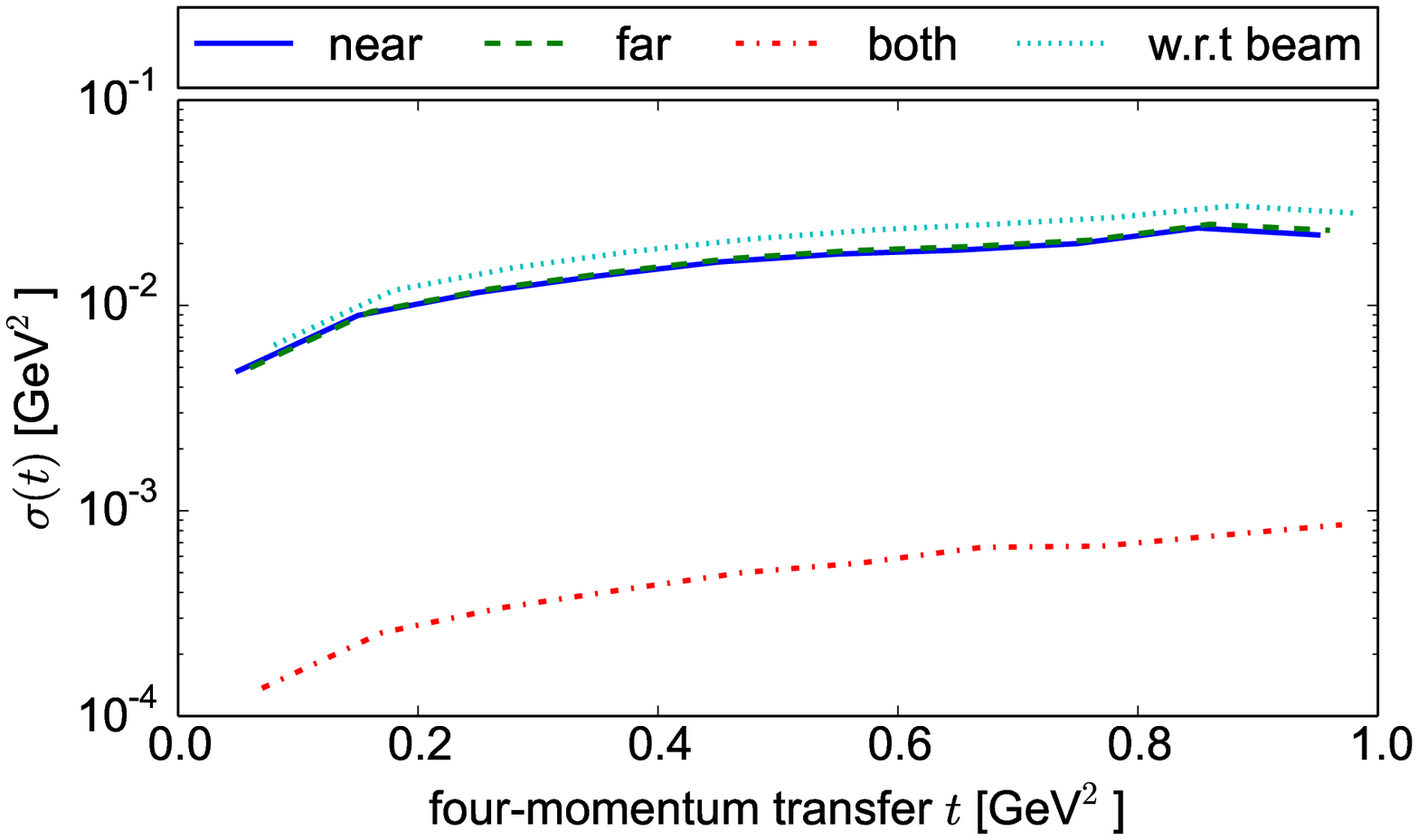}
  \caption{The spread of the reconstruction error of the four-momentum transfer due to a 1 mrad rotation of the detectors.}
  \label{fig:t_rotation_spread}
}

Finally, Figures \ref{fig:phi_rotation_offset} and \ref{fig:phi_rotation_spread} present the error in the case of the azimuthal angle reconstruction.
The pattern is similar as before: the rotations of the near and the far station give opposite offsets and similar spreads.
The simultaneous rotation has a much smaller effect.
The rotation w.r.t. the beam increases the offset.
Figures \ref{fig:phi_t_rotation_offset} and \ref{fig:phi_t_rotation_spread} show how the errors due to the rotation of the near detector depend on the $t$ values.
The events with $t > 0.1\ \text{GeV}^2$ have significantly smaller errors (and smaller spread) than the events with $t < 0.1\ \text{GeV}^2$.

\twofigures{
  \includegraphics[width=\linewidth]{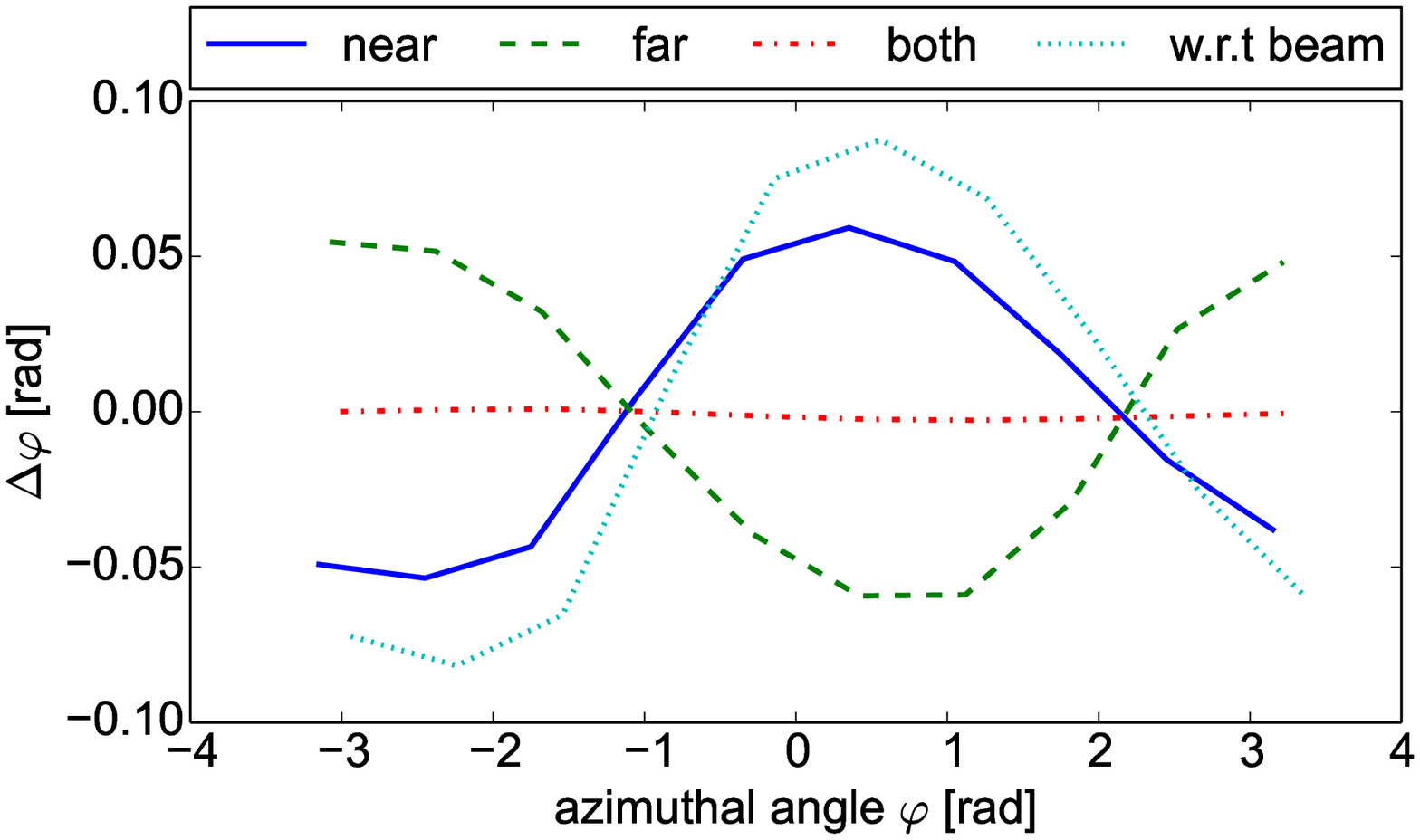}
  \caption{The average reconstruction error of the azimuthal angle due to a 1 mrad rotation of the detectors.}
  \label{fig:phi_rotation_offset}
}
{
  \includegraphics[width=\linewidth]{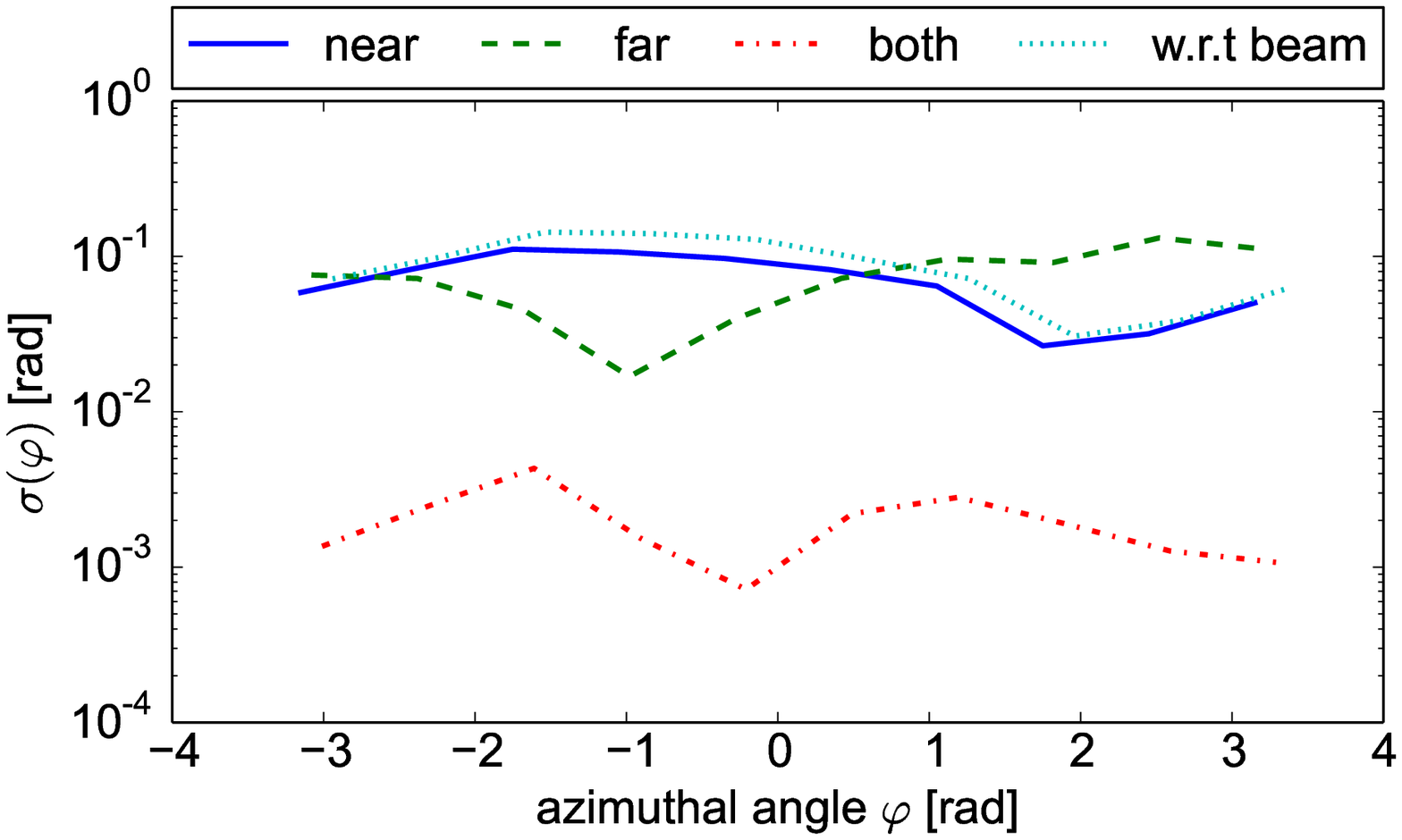}
  \caption{The spread of the reconstruction error of the azimuthal angle due to a 1 mrad rotation of the detectors.}
  \label{fig:phi_rotation_spread}
}

\twofigures{
  \includegraphics[width=\linewidth]{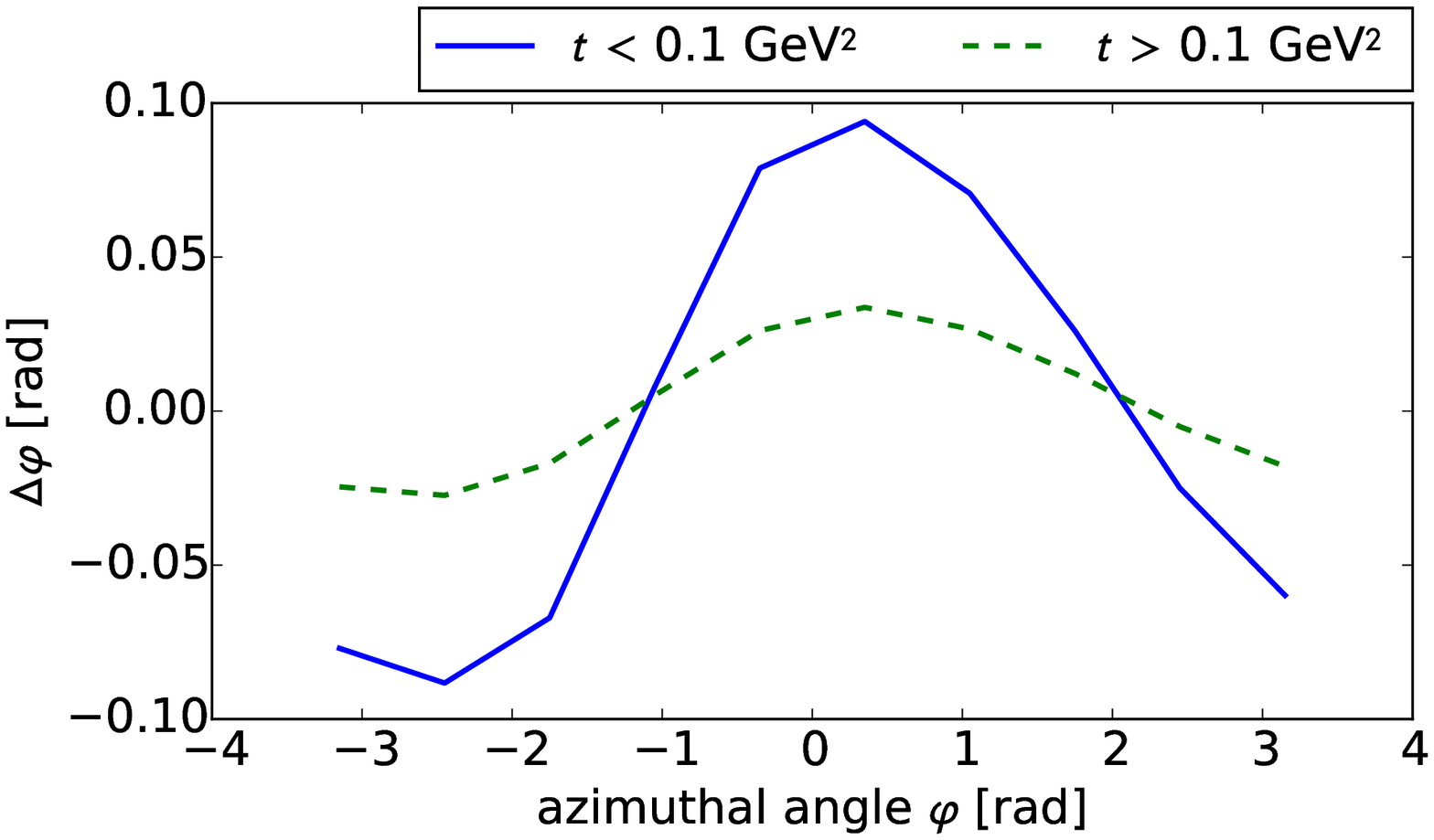}
  \caption{The average reconstruction error of the azimuthal angle due to a 1 mrad rotation of the near detector.}
  \label{fig:phi_t_rotation_offset}
}
{
  \includegraphics[width=\linewidth]{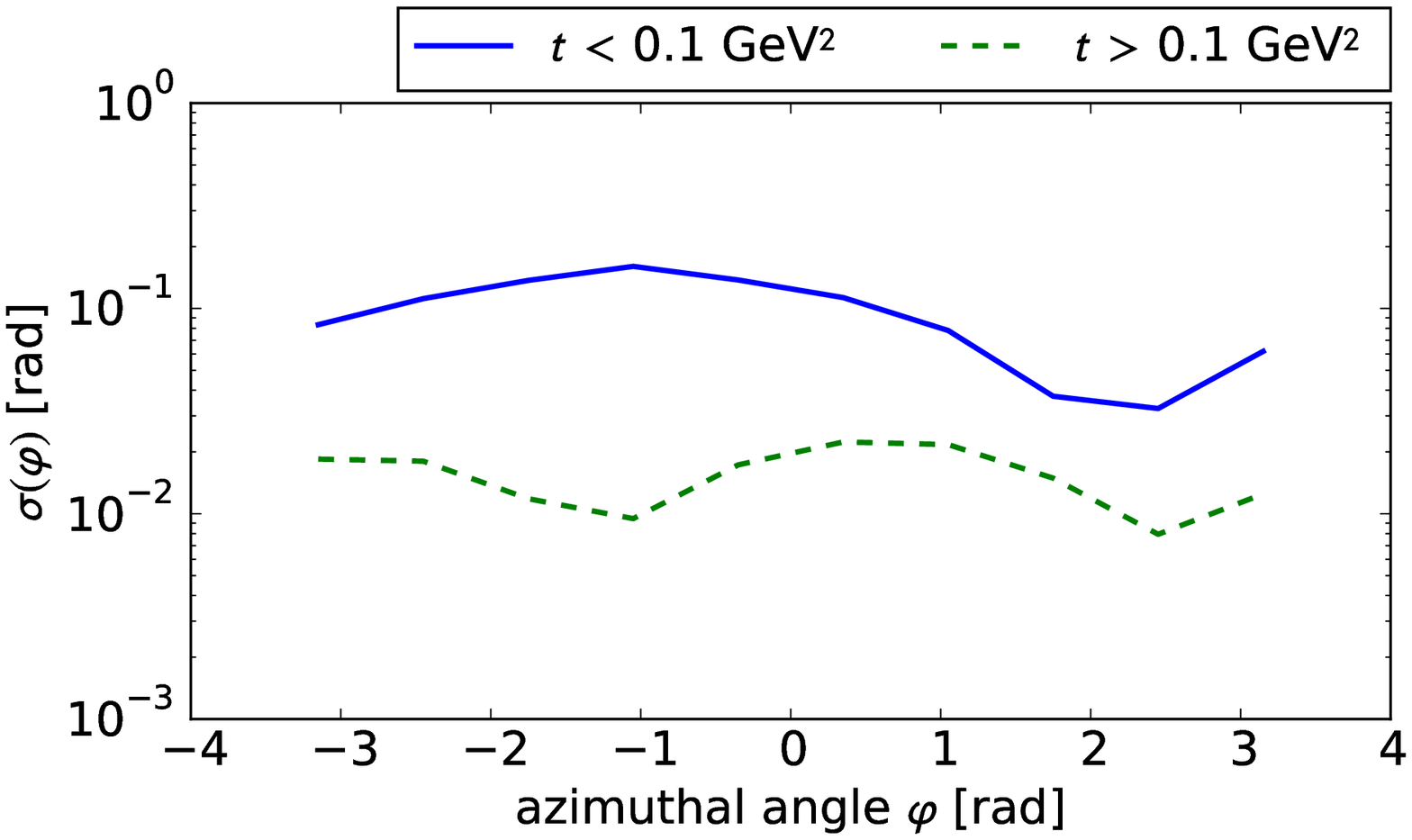}
  \caption{The spread of the reconstruction error of the azimuthal angle due to a 1 mrad rotation of the near detector.}
  \label{fig:phi_t_rotation_spread}
}

\section{Summary and conclusions}

The numerical results presented above provide a foundation upon which the demands for alignment can be examined.
One could consider three types of measurements:
\begin{itemize}
  \item[(a)] \textbf{tag-only} -- no information about the forward
    proton kinematics is used, only the information about its presence in the
    event,
  \item[(b)] \textbf{differential} -- reconstructed kinematics of
    the forward proton is an essential ingredient of the measurement,
  \item[(c)] \textbf{exclusive} -- reconstructed forward proton
    kinematics is needed to discriminate between the signal and the background.
\end{itemize}

An example of a tag-only study is a measurement of the cross section for some diffractive process.
In this case, the only point in which the misalignment can affect the measurement is the estimation of the detector acceptance.
This issue has been discussed in Section 5.
The error due to a misalignment of 100~$\muup$m is largest for the case of soft single diffractive process and amounts to 2 \%.
Assuming that the aim is the cross section measurement with a 10 \% accuracy, the demand on the absolute horizontal alignment precision can be set at 500 $\muup$m.

Examples of differential measurements include the measurements of the $\xi$, $t$ and $\varphi$ distributions, as well as various combinations of these observables with properties of the centrally produced state.
The requirement of a $\xi$ measurement with a precision better than 10 \% implies the absolute horizontal alignment precision of the order of 200 $\muup$m for small $\xi$ values, and 500 $\muup$m for high $\xi$ values.
The relative horizontal precision needs to be more exact and should not exceed 100 $\muup$m.

Measurements of $t$ and $\varphi$ will be dominated by the uncertainty on the relative horizontal and vertical alignments.
The sensitivity of the transverse momentum reconstruction is very high, which leads to the need for the alignment precision of 10 $\muup$m.
On the other hand, this sensitivity provides also a data-driven method for the relative alignment, see \cite{Staszewski:2010fc}.

Exclusive measurements require the kinematics reconstruction not only for the differential measurement, but mainly for background reduction.
For exclusive processes, such as the exclusive production of jets, the kinematics of the centrally produced system (jets) is correlated with the kinematics of the forward protons because of energy-momentum conservation.
For background processes such correlations do not exist or are much weaker.
This makes it possible to select very rare exclusive events.
In order to best exploit these correlations, the proton kinematics must be correctly reconstructed.

In the case of exclusive QCD processes, the most important variable is $\xi$, which can be correlated to the mass and rapidity of the produced jet system.
Since exclusive processes are rare, one expects mainly events with small $\xi$ values because the distribution is falling as $1/\xi$ with increasing $\xi$.
This leads to the requirement on the absolute horizontal alignment precision better than 100 $\muup$m and the relative horizontal alignment precision of 25 $\muup$m.
For QED processes, such as the exclusive production of lepton pairs, also $t$ and $\varphi$ can be used, because of the possibility of a precise $p_T$ measurement of the lepton pair in the central detector.
Here, also the relative misalignment plays a role, and the requirement is similar as for the $t$ and $\varphi$ measurement: a precision of 10~$\muup$m.

\section*{Acknowledgements}

We gratefully acknowledge Peter Bussey for many stimulating discussions on detector alignment and machine optics.

This work was supported in part by the Polish National Science Centre grant:\\UMO-2012/05/B/ST2/02480.



\bibliographystyle{model1-num-names}

\bibliography{sample}

\end{document}